\documentclass[11pt]{article}

\usepackage[final]{acl}

\usepackage{times}
\usepackage{latexsym}
\usepackage[T1]{fontenc}
\usepackage[utf8]{inputenc}
\usepackage{microtype}
\usepackage{inconsolata}
\usepackage{booktabs}
\usepackage{amsfonts}
\usepackage{amsmath}
\usepackage{graphicx}
\usepackage{tabularx}
\usepackage{multirow}
\usepackage{placeins}
\usepackage{stfloats}
\usepackage{float}
\usepackage{xcolor, soul,todonotes} 
\definecolor{kmycolor}{rgb}{0.858, 0.188, 0.478}

\newcommand{\rev}[1]{#1}

\newcommand{\revb}[1]{#1}

\title{When Models Edit Too Much: On the Fidelity of Minimal Code Edits}
\author{
  Tongyao Zhu\thanks{Equal contribution} \qquad Wei Hern Lim$^*$ \qquad Min-Yen Kan\thanks{Corresponding author} \\
  National University of Singapore \\
  \texttt{\{tongyao.zhu, limweihern\}@u.nus.edu} \qquad \texttt{knmnyn@nus.edu.sg}
}

\begin{document}

\maketitle

\begin{abstract}
\looseness=-1 Large language models (LLMs) are increasingly used to edit existing code, but correctness alone is not enough: useful repairs should also be minimal, reviewable, and faithful to the original implementation. We study over-editing, the tendency of a model to rewrite code beyond what is required to fix a bug. We construct an evaluation framework from 400 BigCodeBench problems by injecting controlled AST-level corruptions into reference solutions, giving each repair task a known minimal patch. Across frontier LLMs, over-editing is widespread even among strong models like \rev{GPT-5.5}: high Pass@1 can coexist with unnecessarily large edits and added cognitive complexity. A preservation instruction substantially reduces this behavior, lowering average excess Levenshtein distance from 0.195 to 0.131, reducing added cognitive complexity by 26.6\%, and increasing Pass@1 by 2.3 points. However, these gains do not simply follow from a larger reasoning budget or larger models. We next ask whether minimal editing can be learned directly during post-training. We observe that supervised fine-tuning overfits to seen corruption patterns, whereas reinforcement learning gives the \rev{best} out-of-domain edit-fidelity and performance-retention trade-off. These results position edit fidelity as a distinct axis of code-repair quality and show that it can be measured and learned.\rev{\footnote{Code: \url{https://github.com/nreHieW/over-editing}}}
\end{abstract}

\section{Introduction}
Software engineering is one of the main practical use cases for frontier LLMs, including systems explicitly optimized for coding assistance \citep{openai2025_gpt5_systemcard,anthropic2025_claude_sonnet4}. In actual development workflows, developers \rev{need more than} patches that execute correctly; they need patches that are easy to review, preserve the original intent of the code, and avoid introducing new complexity. This requirement is especially important in brownfield settings, where the model edits an existing implementation rather than generating a new one from scratch. \rev{In such maintenance work, keeping patches minimal reduces review burden, avoids diff churn, preserves implicit design choices, and lowers the risk of regressions outside existing tests.}

\begin{figure}[t]
  \centering
  \includegraphics[width=\columnwidth]{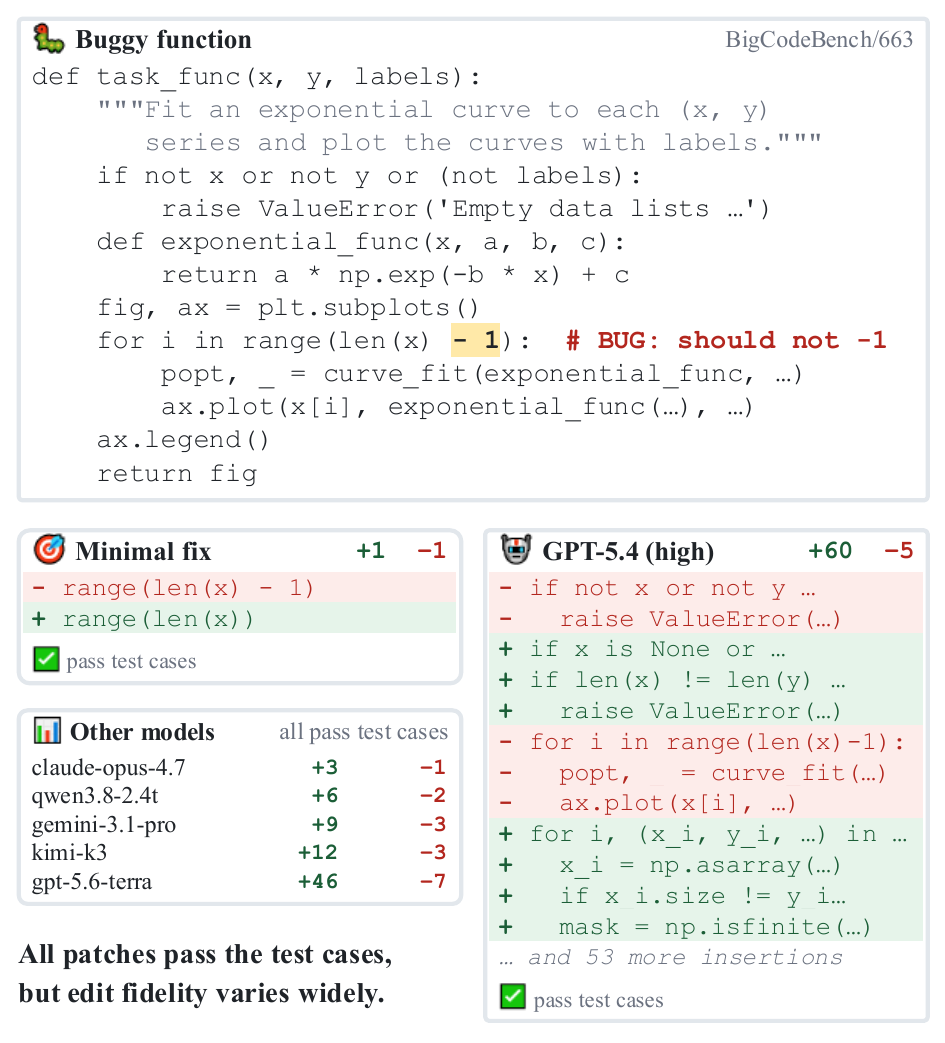}
  \caption{
  A one-line off-by-one bug in a BigCodeBench function, repaired by a minimal
  patch and by six frontier models.
  Every patch passes the task's five tests: the minimal fix changes a single
  line, while GPT-5.4 adds 60 lines of input validation, dtype coercion, NaN
  masking, and resampling that no test requires. \emph{Other models} gives
  median changes over five completions on the identical input,
  for models whose completions all pass the test cases.}\vspace{-3mm}
  \label{fig:over-editing-example}
\end{figure}

A model can successfully fix a bug while still rewriting much more of the function than necessary. We refer to this behavior as \emph{over-editing}: a functionally correct repair that changes more code than the minimal fix requires. Figure~\ref{fig:over-editing-example} is illustrative: the buggy function requires only a one-line boundary correction, and we verify that this single-line patch passes all five of the task's tests. GPT-5.4 instead deletes five lines and inserts 60, spanning input validation, dtype coercion, NaN masking, and curve resampling. The output passes the same five tests, so none of that additional code is required by the specification. \rev{Running the identical input through 20 frontier models indicates that this is a property of the model rather than of the task: among models whose five completions all pass the test cases, median patch size ranges from 2 to 60 inserted lines, and the most recent releases are not uniformly more conservative than the models they replace.}

Existing coding benchmarks primarily measure functional success, through Pass@1 or Pass@\(k\) on tasks \citep{chen2021evaluatinglargelanguagemodels,liu2023codegeneratedchatgptreally,austin2021programsynthesislargelanguage,li2023tacotopicsalgorithmiccode,zhuo2025bigcodebenchbenchmarkingcodegeneration,jain2024livecodebenchholisticcontaminationfree}. Repository-level and direct-editing benchmarks broaden the setting, but still mostly evaluate whether tests pass or issues are resolved \citep{jimenez2024swebenchlanguagemodelsresolve,Chowdhury2024swebenchverified,Zheng2024HumanEvoAEB,He2025SWEPerfCLD,guo2025codeeditorbenchevaluatingcodeediting,tian2024debugbenchevaluatingdebuggingcapability,gauthier2024aider,fu2025corecodebenchconfigurablemultiscenariorepositorylevel}. Concurrent work PDB \citep{chai2026precise} similarly highlights that passing tests does not guarantee precise, targeted repairs. We likewise aim to make \emph{edit fidelity} measurable in its own right: not only whether a model fixes a bug, but whether it preserves the surrounding implementation when the required repair is local.

We construct a controlled evaluation framework from 400 BigCodeBench problems \citep{zhuo2025bigcodebenchbenchmarkingcodegeneration}. For each reference solution, we inject \rev{one or two} localized AST-level corruptions and retain the example only if the corrupted program fails the original tests. This gives each task a known minimal repair: the reversal of the injected corruptions. We evaluate model outputs with Pass@1 for functional success, normalized token-level Levenshtein distance for edit size, and added cognitive complexity for structural overhead. These metrics evaluate whether the model fixes a bug with minimal edits that are easier to understand and review. \rev{As the minimal repair is known by construction, the framework isolates over-editing as a failure: a model unnecessarily rewriting functional code on a localized repair task. Our scope is deliberately local code repair; open-ended generation, architectural refactoring, and feature extension, where broader rewrites are intentional, fall outside it.}

Our results show that edit fidelity is a distinct axis of code-repair quality:
\textbf{(1) Over-editing is widespread:} several frontier models, such as GPT-5.5, achieve competitive Pass@1 while still making unnecessarily large edits.
\textbf{(2) Prompting helps:} a single preservation instruction reduces aggregate frontier excess Levenshtein distance from \(0.195\) to \(0.131\), lowers added cognitive complexity by \(26.6\%\), and improves Pass@1 by \(2.3\) points.
\textbf{(3) Reasoning and scale are insufficient:} reasoning effects are model-specific, and larger models do not monotonically produce smaller or simpler passing repairs.
\textbf{(4) Minimal editing can be learned:} supervised fine-tuning overfits to seen corruption patterns, while reinforcement learning reaches \(0.782\) out-of-domain Pass@1 with \(0.050\) excess Levenshtein distance and does not degrade general coding ability.
Qualitative analysis further suggests that over-editing is often a granularity mismatch: the model may identify the bug but rewrite data flow, add defensive checks, or change nearby behavior instead of making the local repair. Together, these findings show that over-editing is widespread, measurable, and reducible through prompting and post-training.

\section{Related Work}
\textbf{Correctness-centered code evaluation.}
Coding LLMs are usually evaluated by functional success, such as Pass@1 or Pass@\(k\), on executable generation benchmarks \citep{chen2021evaluatinglargelanguagemodels,liu2023codegeneratedchatgptreally,austin2021programsynthesislargelanguage,li2023tacotopicsalgorithmiccode,ni2023l2cevalevaluatinglanguagetocodegeneration,zhuo2025bigcodebenchbenchmarkingcodegeneration,jain2024livecodebenchholisticcontaminationfree}. Repository-level and debugging benchmarks broaden the setting to issue resolution or repair-like tasks \citep{jimenez2024swebenchlanguagemodelsresolve,Chowdhury2024swebenchverified,Zheng2024HumanEvoAEB,He2025SWEPerfCLD,tian2024debugbenchevaluatingdebuggingcapability,fu2025corecodebenchconfigurablemultiscenariorepositorylevel}, but task success alone does not show how much of an existing implementation a model rewrites. This concern is aligned with classic automated program repair work distinguishing plausible patches from correct, maintainable, or reviewable ones \citep{qi2015plausibility,liu2021criticalreview}.

\textbf{Editing and minimal repair.}
Early function-level repair benchmarks include HumanEvalFix \citep{muennighoff2023octopack}. More recent instructed-editing benchmarks test whether models can modify existing code from natural-language requests \citep{cassano2024canitedit,guo2025codeeditorbenchevaluatingcodeediting,chi2025editbench}. CanItEdit also reports ExcessCode for superfluous changed lines, and benchmark audits show that weak test oracles can miss extraneous edits \citep{cassano2024canitedit,ebrahimi2026editverify}. Concurrent work PDB \citep{chai2026precise} injects verified bugs into correct programs, including BigCodeBench solutions, and measures edit-level precision and bug-level recall. Complementary repair systems explicitly target smaller or more faithful patches: CREF \citep{yang2024cref} measures patch precision in tutoring, AdaPatcher \citep{dai2025adapatcher} uses localization and preference learning, and PAFT \citep{yang2026paft} trains preservation-aware minimal-edit repair models. Concurrent work PRepair \citep{ke2026prepair} studies edit-aware RL for precise repair.  In contrast, our focus is a controlled BigCodeBench evaluation with known minimal reversals, spanning frontier models, prompts, reasoning variants, and post-training methods.

\textbf{Similarity metrics and constraints.}
Reference-similarity metrics such as CodeBLEU are widely used for generated code \citep{Evtikhiev_2023,ren2020codebleumethodautomaticevaluation}, but they do not directly ask whether a repair changed more than necessary relative to a buggy input and its minimal fix. Reasoning models also raise a code-specific constraint-following question: although reasoning often improves coding and instruction-following performance \citep{zhou2023instructionfollowingevaluationlargelanguage}, recent work shows that stronger reasoning can interact poorly with explicit constraints \citep{fu2025scalingreasoninglosingcontrol,li2025thinkingfailspitfallsreasoning,wen2024benchmarkingcomplexinstructionfollowingmultiple}. We study that interaction for faithful code repair.

\section{Evaluating Over-Edit Behavior}
We organize our study around four questions:
\begin{itemize}
    \item \textbf{RQ1:} Do strong coding models over-edit even when their repairs pass tests?
    \item \textbf{RQ2:} Can prompting reduce over-editing?
    \item \textbf{RQ3:} How do factors like reasoning or model size interact with edit fidelity?
    \item \textbf{RQ4:} When and how do models over-edit?
\end{itemize}

\begin{figure*}[t]
  \centering
  \includegraphics[width=\textwidth]{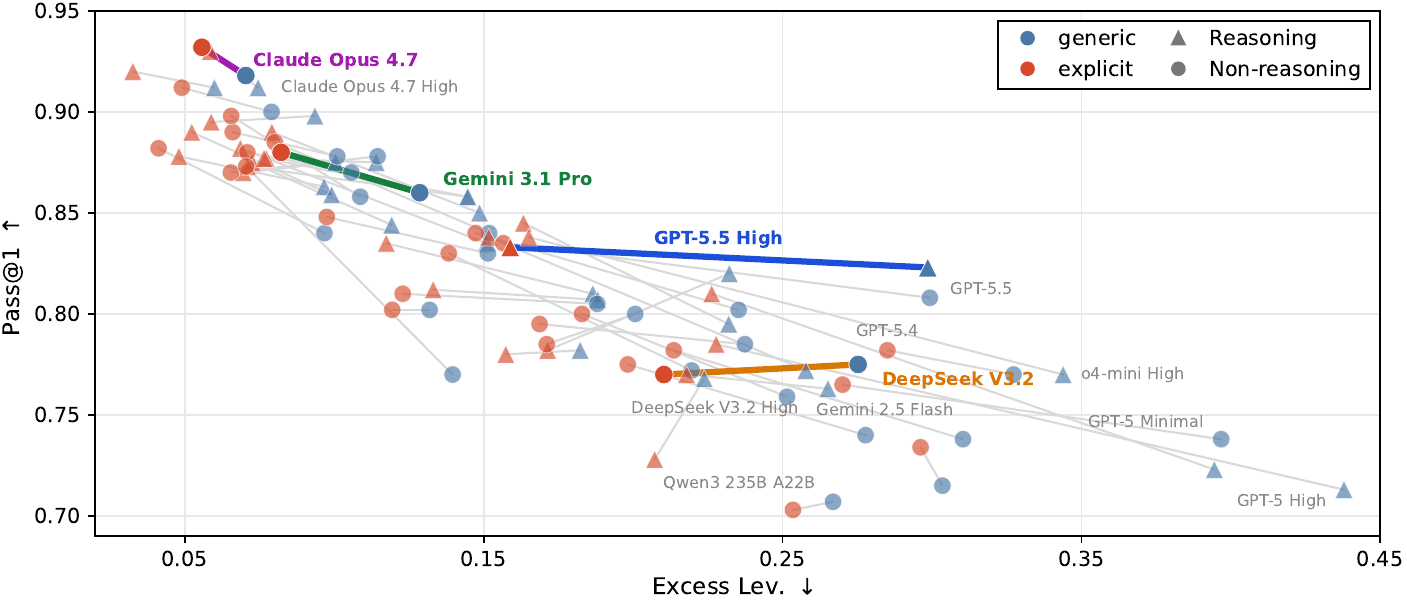}
  \caption{Frontier model performance under generic and preservation prompts. Points show individual model-prompt settings; connected pairs show how the same model changes when preservation is made explicit. Triangles denote reasoning variants and circles denote non-reasoning variants; model names are shown where space permits. \rev{Models discussed in the text are highlighted in color; all other settings are shown faded.}}
  \label{fig:frontier_per_model_prompt_effect}
\end{figure*}

\subsection{Setup}
\label{sec:eval-setup}
\paragraph{Benchmark Construction.} We sample 400 problems from BigCodeBench \citep{zhuo2025bigcodebenchbenchmarkingcodegeneration}, which provides diverse Python tasks together with reference solutions and executable tests. For each problem, we start from the reference implementation, inject \rev{one or two} controlled AST-level corruptions taken from a predefined list (Appendix~\ref{sec:AppendixCorruptions}), and retain the example only if the corrupted solution fails the original tests. The resulting task is therefore a function-level repair problem with a known minimal target patch (the reversal of the corruptions applied).

We use manual corruption rather than LLM-generated bugs for two reasons. First, it gives finer control over the locality and semantic type of the bug, which is important when the quantity of interest is \emph{how much} the model \rev{over-edits}.
Second, it keeps the evaluation interpretable: when a repair is unnecessarily large, we can meaningfully compare it against the small set of intended local changes. By starting from a correct reference solution and injecting localized corruptions, we obtain tasks where the intended repair is small by design. This is the key advantage of the benchmark: because the bug is injected by construction, the minimal repair is known rather than inferred.

\rev{
\paragraph{Benchmark statistics.} The benchmark consists of short functions with an average of 10.4 executable lines, up to a maximum of 34. The gold repairs are small by construction: 50.2\% require a single token edit, 91.8\% at most two, and none touches more than two lines. Of the 400 tasks, 232 carry one injected corruption and 168 carry two, giving a total of 568 corruption applications. Semantically, predicate/control-flow bugs account for 43.0\%, computation/value for 29.9\%, boundary/iteration for 18.7\%, and API/data semantics for 8.5\% (full distributions in Appendix~\ref{sec:AppendixBenchmarkStats}).}

\paragraph{Metrics.} We evaluate each repair along three axes: whether it works, how much it changes relative to the known minimal repair, and whether it adds structural complexity. Unless stated otherwise, edit-fidelity metrics use passing repairs.

\textbf{Functional success.} Pass@1 is the fraction of model outputs that pass all tests for the corresponding problem. It captures the basic repair objective, but not edit fidelity: a model can pass the tests while rewriting far more code than the bug requires. We therefore interpret the following edit-size metrics alongside Pass@1.

\textbf{Token-level edit distance.} Let \(C\) be the corrupted solution, \(G\) the gold repair, and \(M\) the model output. We compare extracted function bodies after removing comments and formatting-only artifacts, tokenize Python code, and compute Levenshtein distance over tokens rather than characters. For two solutions \(X,Y\), \(d(X,Y)\) is the token Levenshtein distance normalized by the larger token count of the two extracted bodies. This treats keywords, identifiers, operators, and punctuation as code units and avoids identifier-length artifacts.

The key quantity is the excess normalized edit distance relative to the known repair:
\[
D_{\text{gold}} = d(G, C), \qquad D_{\text{model}} = d(M, C),
\]
\[
E_{\text{Lev}}(M) = D_{\text{model}} - D_{\text{gold}}.
\]
\(D_{\text{gold}}\) is the size of the minimal patch that reverses the injected corruption, while \(D_{\text{model}}\) is the size of the model's patch from the same corrupted input. We interpret this excess-normalized value \rev{as follows}:
Any value larger than zero means that the model changes more code than the gold repair, and large positive values indicate excess editing.

We prefer token-level edit distance to higher-level n-gram metrics such as CodeBLEU~\citep{ren2020codebleumethodautomaticevaluation} because our corruptions are intentionally local: n-gram overlap can reward superficial stylistic
similarity even when the patch is unnecessarily large, whereas edit distance on tokens tracks whether
the model disturbed the implementation beyond reversing the bug. Appendix~\ref{sec:AppendixMetricSanity} shows high-disagreement cases.

\textbf{Added cognitive complexity.} Edit size does not capture every form of over-editing: a short patch can still introduce extra branches, nesting, or restructuring that make the result harder to review. We therefore report added cognitive complexity~\citep{campbell2021cognitivecomplexity}, measured with a Python AST visitor as the cognitive complexity of \(M\) minus that of \(G\). As our corruptions are local changes rather than structural rewrites, any complexity increase beyond the gold repair reflects unnecessary overhead for human reviewers.

\rev{\paragraph{Human validation of the metrics.} We validate that the metrics reflect human perception of code edits. Three annotators with five to ten years of development experience compared 100 blinded pairs of passing repairs of the same corrupted programs, choosing the patch that was easier to review and the one more faithful to the original. Excess Levenshtein distance matches the human majority in 94.8\% of decided cases for reviewability (Cohen's $\kappa=0.897$) and 96.9\% for faithfulness ($\kappa=0.939$); added cognitive complexity agrees moderately ($\kappa=0.455$ and $0.386$ for reviewability and faithfulness, respectively). Annotators agree strongly with one another (Fleiss' $\kappa=0.690$ and $0.850$). We further perform a blinded audit of 100 high-excess passing repairs and find genuinely unnecessary edits in 82.3\% of determinate cases (95\% CI 73.5--88.6\%), while the rest are valid alternative fixes (Appendix~\ref{sec:AppendixHumanStudies}).}

\paragraph{Prompt Setup.} We evaluate models in two conditions. In the \textbf{generic} setting, the model is simply asked to fix the bug. In the \textbf{explicit} setting, we \rev{append a preservation clause to the same request}, asking the model to preserve the original code and modify only what is necessary (full prompt in Appendix~\ref{sec:AppendixPrompt}). This comparison tests whether over-editing is partly reducible through prompting alone, without changing the underlying task or test suite.  Appendix~\ref{sec:AppendixEvalDetails} gives details.

\subsection{Results}
\label{sec:eval-results}
\paragraph{Frontier LLMs over-edit by default.}
\looseness=-1 Figure~\ref{fig:frontier_per_model_prompt_effect} summarizes the frontier results, with full numbers in Appendix~\ref{sec:AppendixFrontierTables}. In the generic setting, correctness and edit fidelity separate: many strong models can repair the program while still changing substantially more code than the known minimal reversal. This behavior is not simply a symptom of weak coding ability. Some of the same models that pass many tasks also add validation, restructure intermediate computations, or change surrounding control flow. The default repair policy therefore appears closer to ``produce a robust solution'' than to ``recover the smallest local patch.'' Claude Opus 4.7 \citep{anthropic2026claudeopus} \revb{(Figure~\ref{fig:frontier_per_model_prompt_effect}, purple)} gives the \rev{best} non-reasoning trade-off, showing that high correctness and small edits can coexist, but models such as GPT-5.5 \citep{openai2026_gpt55} \revb{(blue)}, DeepSeek \citep{deepseek_v31_2025,deepseekai2025deepseekr1incentivizingreasoningcapability,deepseekai2025deepseekv3technicalreport,deepseekai2025deepseekv32} \revb{(orange)}, and Gemini \citep{google2025_gemini25_flash,google2026_gemini31_pro} \revb{(green)} variants show that Pass@1 alone can hide large excess edits\revb{: GPT-5.5 High reaches Pass@1 $0.823$ yet its excess distance is over four times Opus 4.7's}.

\begin{figure}[t]
  \centering
  \includegraphics[width=\columnwidth]{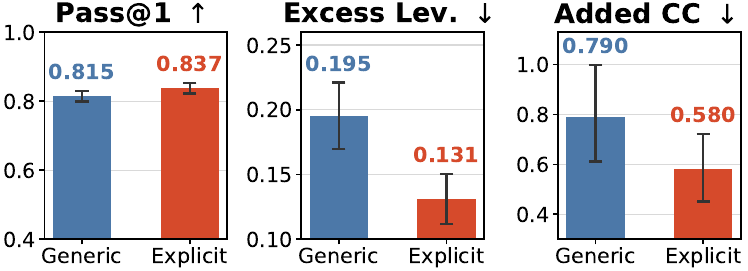}
  \caption{Aggregate effect of preservation prompting on frontier models\revb{: means over the 50 settings of Table~\ref{tab:frontier_full}; whiskers give bootstrap 95\% CIs}. Making preservation explicit improves Pass@1 and reduces both excess edit distance and added cognitive complexity \revb{(all significant under matched-pair tests, Appendix~\ref{sec:AppendixRobustness})}.}
  \label{fig:frontier_prompt_aggregate}
\end{figure}

\paragraph{Explicit prompting improves edit fidelity.}
\looseness=-1 We next investigate whether adding an explicit instruction to minimally edit the code improves performance. \revb{The generic request ends with \textit{``Fix and complete my function.''}, while the explicit one adds \textit{``\ldots but keep as much of the original code as possible''} (full prompt in Appendix~\ref{sec:AppendixPrompt}).} \revb{Figure~\ref{fig:frontier_per_model_prompt_effect} shows the per-model effect: the clause markedly reduces over-editing, shifting all 50 settings left, and slightly improves quality, raising Pass@1 in 40 of 50. The heaviest over-editors move furthest --- GPT-5.5 High (blue) nearly halves its excess distance ($0.299$ to $0.159$) --- while the already-faithful Opus 4.7 (purple) barely moves. In aggregate (Figure~\ref{fig:frontier_prompt_aggregate}), excess Levenshtein distance drops from $0.195$ to $0.131$, added cognitive complexity falls by 26.6\% (matched-pair signed-rank $p<10^{-4}$ for both), and Pass@1 rises by 2.3 percentage points (paired bootstrap 95\% CI $[+1.49,+3.05]$); the gain persists under repeated temperature-1 sampling and paraphrased prompt variants (Appendix~\ref{sec:AppendixRobustness}).} Thus, the prompt appears to select a different latent repair mode: the model treats the existing implementation as evidence to preserve, rather than as a reference to improve on.
We additionally verify that many smaller open-weight models behave similarly \rev{(Appendix~\ref{sec:AppendixOpenWeightPrompt}):} \revb{the same clause} improves average Pass@1 from \(0.788\) to \(0.828\) while reducing excess Levenshtein distance from \(0.176\) to \(0.121\). Overall, this suggests that over-editing is a prevalent yet steerable behavior across all major LLMs.

\begin{figure}[t]
  \centering
  \includegraphics[width=\columnwidth]{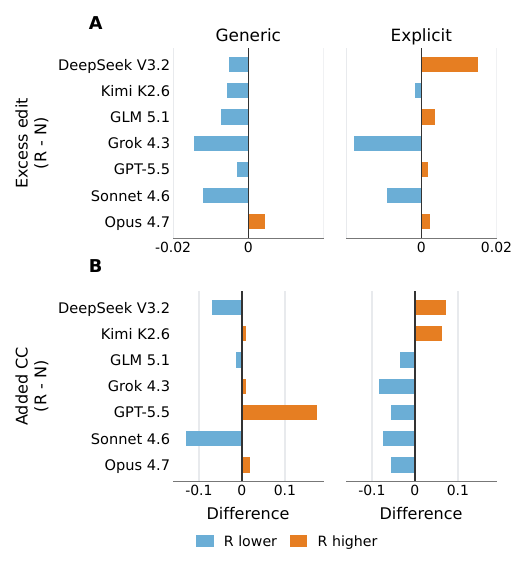}
  \caption{Effect of reasoning on edit fidelity within matched frontier model families. Bars compare reasoning and non-reasoning variants under each prompt; negative values indicate smaller edits or lower added complexity from reasoning.}

  \label{fig:frontier_reasoning_aggregate}
\end{figure}

\paragraph{Reasoning does not necessarily reduce over-editing.}
\looseness=-1 In Figure~\ref{fig:frontier_reasoning_aggregate}, we compare matched frontier reasoning and non-reasoning variants under both prompts. Each bar is the reasoning value minus the non-reasoning value on the jointly passing subset, so negative bars mean reasoning yields a more faithful repair. The pattern is model-specific rather than monotonic. Under the generic prompt, reasoning slightly reduces excess edits for most models shown, while Claude Opus 4.7 is narrowly reversed. Added cognitive complexity is more mixed: Sonnet 4.6 and DeepSeek V3.2 reduce complexity with reasoning, but GPT-5.5 increases it substantially. Under the explicit preservation prompt, Grok~4.3 and Sonnet~4.6 become more faithful on both metrics, whereas DeepSeek~V3.2 shows more excess edits and added complexity with reasoning. Thus, reasoning alone is not a universally reliable mitigation \rev{for} over-editing. 

\begin{figure*}[t]
  \centering
  \includegraphics[width=\textwidth]{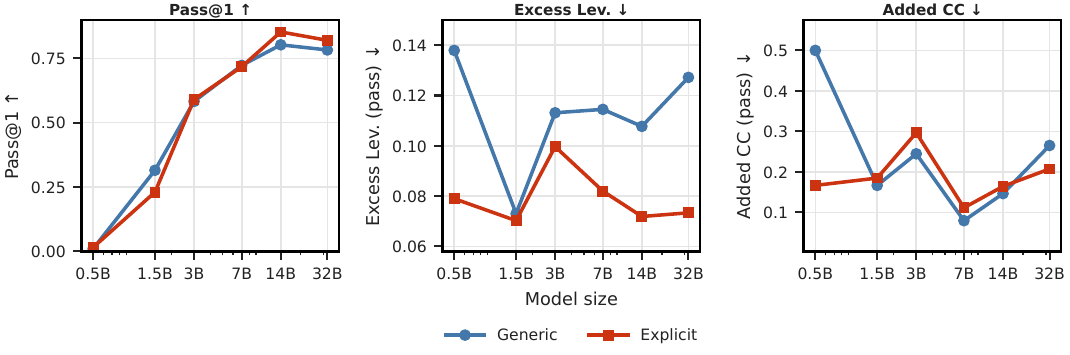}
  \caption{Performance of Qwen2.5-Coder-Instruct across model sizes. Scaling improves Pass@1, but edit fidelity does not improve monotonically; preservation prompting consistently reduces excess edits.}
  \label{fig:qwen25_size_sweep}
\end{figure*}

\paragraph{Size scaling does not monotonically reduce over-editing.}
\looseness=-1 We further investigate whether over-editing is linked to model size. We select the Qwen2.5-Coder-Instruct series \citep{hui2024qwen25coder} because it provides model sizes from 0.5B to 32B. Figure~\ref{fig:qwen25_size_sweep} reports Pass@1 over all 400 examples, while the edit-fidelity panels are computed only over repairs that pass the tests. This avoids giving small models credit for failed no-edit outputs. For the smallest models, the edit-fidelity points should therefore be read as conditional averages over the few successful fixes, not as evidence that the model is reliably faithful. Pass@1 generally increases as the model grows larger, as expected, but excess Levenshtein distance and added cognitive complexity do not improve monotonically among the successful repairs. Scale therefore moves models out of the regime where they simply fail to repair, but it does not guarantee that the passing repairs become more local\revb{: under the generic prompt, excess distance rises again from $0.108$ at 14B to $0.127$ at 32B}. Larger models may have stronger task-solving priors, but those priors still need to be aimed at preservation.

\paragraph{What bugs trigger over-editing?}
\looseness=-1 We then ask which injected bugs push models away from local repair.  Table~\ref{tab:corruption_overedit} shows where over-editing is concentrated. We observe that \rev{list-related operations} (like slicing and indexing) and conditionals trigger \rev{the} most over-editing. We think that the important pattern is ambiguity: a one-token bug can look like evidence of missing preconditions, unsafe indexing, or unstable control flow. The model often treats the existing implementation as suspect, even when the intended repair is small. Moreover, this is not simply a difficulty effect: high Pass@1 can coexist with large excess edits, \revb{as with slice bounds with the highest Pass@1 ($0.874$) but the largest excess ($0.353$)}. This suggests that over-editing itself is an important dimension that should be evaluated \rev{alongside} editing correctness.

\begin{table}[t]
  \centering
  \small
  \caption{\looseness=-1 Corruption types most associated with over-editing. Boundary and control-flow bugs produce high excess edit distance or added complexity even when repairs pass.}
  \label{tab:corruption_overedit}
  \setlength{\tabcolsep}{3pt}
  \newcommand{\correxample}[2]{#1\newline\hspace*{1em}{\footnotesize\texttt{#2}}}
  \begin{tabular}{@{}>{\raggedright\arraybackslash}p{0.37\columnwidth}ccc@{}}
    \toprule
    Corruption\newline\hspace*{1em}{\footnotesize Example} & Pass@1 & Excess Lev. & Added CC \\
    \midrule
    \correxample{Slice bounds}{x[a:b]} & 0.874 & 0.353 & 1.25 \\
    \correxample{List indexing}{list[i]} & 0.780 & 0.244 & 1.14 \\
    \correxample{Comparison ops.}{x < y} & 0.799 & 0.210 & 0.78 \\
    \correxample{Sort order}{reverse=True} & 0.709 & 0.239 & 0.80 \\
    \correxample{Conditional inv.}{if cond:} & 0.778 & 0.207 & 1.46 \\
    \correxample{Edge-case guards}{if not list:} & 0.791 & 0.205 & 0.94 \\
    \bottomrule
  \end{tabular}
\end{table}

\paragraph{How do models over-edit?}
\looseness=-1 We next look at the behavior when the model over-edits. \rev{We ask GPT-5.5 to design a taxonomy of over-editing behavior, freeze its categories into an annotation codebook, and label every high-excess passing repair (excess Levenshtein distance $\geq 0.5$; $n=530$) from the five frontier models. Appendix~\ref{sec:AppendixTaxonomyValidation} validates the labels: five independent models apply the categories consistently, the shares change little across data subsets and under re-annotation, and near-minimal repairs almost never receive a label.} Table~\ref{tab:overedit_issue_types} suggests that over-editing is usually a change in repair granularity, not just extra style edits (Appendix~\ref{sec:AppendixOvereditExamples} gives representative examples). The model may replace the data path, wrap the function in defensive checks, or shift the output contract. These moves can be reasonable in open-ended programming, but they are misaligned when the input is already a near-correct program, as they introduce additional burden for reviewers, and make the code \rev{take} much longer to read.


Overall, over-editing looks like a task-framing mismatch. The model behaves as if asked to deliver robust code, while we measure whether it recovers the original intent. Preservation prompting helps because it corrects that frame by \rev{explicitly} instructing the model \rev{to preserve the original code}.

\begin{table}[t]
  \centering
  \small
  \setlength{\tabcolsep}{3pt}
  \caption{\rev{Common over-editing patterns across all $530$ high-excess passing repairs from five frontier models. Categories are multi-label, so a single repair may appear in multiple rows.}}
  \label{tab:overedit_issue_types}  \begin{tabular}{@{}>{\raggedright\arraybackslash}p{0.34\columnwidth}r>{\raggedright\arraybackslash}p{0.46\columnwidth}@{}}
    \toprule
    Issue & Share & Meaning \\
    \midrule
    Defensive generalization & 64.2\% & Adds broad validation, checks, or fallbacks. \\
    Data-flow rewrite & 63.2\% & Re-solves the task instead of locally patching it. \\
    Contract drift & 34.7\% & Changes outputs, side effects, or mutation. \\
    Feature accretion & 23.6\% & Adds plotting or reporting polish. \\
    Dependency fallback & 3.2\% & Adds backup resources or hardcoded data. \\
    \bottomrule
  \end{tabular}
\end{table}

\section{Learning Minimal Editing}
We have earlier shown that models can be steered toward faithful edits at inference time. We next ask whether that behavior can be made durable through training. We study this question with Qwen3 Instruct models because of their strong coding ability with open weights, using the same controlled corruption framework as in Section~\ref{sec:eval-setup}.

\subsection{\revb{Setup}}
\label{sec:Training}
We fine-tune Qwen3-4B-Instruct-2507 \citep{yang2025qwen3technicalreport} on corrupted code from DeepCoder \citep{luo2025deepcoder} with \rev{4{,}141 randomly drawn training examples and 400 test instances}. 
Unlike in the BigCodeBench evaluation split, where each example receives \rev{one or two} corruptions, we apply between one and ten corruptions \rev{per training instance} and \rev{keep} only corrupted programs that fail the tests.

During training, we use the same corruption types as those in Section~\ref{sec:eval-setup}. During evaluation, we \rev{use} both the \emph{in-domain} corruption types \rev{(the same as in training)} \rev{and} another 20 \emph{out-of-domain} \rev{(OOD)} corruptions (listed in Appendix~\ref{sec:AppendixTrainingCorruptions}) to test \rev{generalization}. 

\looseness=-1 We compare SFT, \rev{rejection-sampled} SFT (rSFT), Direct Preference Optimization (DPO) \citep{rafailov2024directpreferenceoptimizationlanguage}, and RL. SFT trains on programmatic minimal repairs. For rSFT and DPO, we generate eight candidate repairs per training sample, keep passing candidates, and rank them by Levenshtein distance to the reference repair; rSFT uses the three smallest-edit candidates, while DPO prefers the smallest-edit candidate over the largest-edit one. These methods are trained with LlamaFactory \citep{zheng2024llamafactory}. RL samples \(K=16\) repairs per corrupted program and scores them with execution feedback plus edit minimality. We use a GRPO-style group-relative RL objective implemented in PRIME-RL \citep{primeintellect2025prime-rl}. \rev{Appendices~\ref{sec:AppendixTrainingRobustness} and~\ref{sec:AppendixRolloutAblation} show that our conclusions hold with exactly one corruption in both training and evaluation, and with RL rollout budgets matched to the 8 candidates of rSFT/DPO.}

Let \(C\) be the corrupted program, \(G\) the gold repair, and \(M\) a sampled repair. Failed or unparsable repairs receive \(r(M)=-0.2\). For passing repairs, the default RL reward is 
\[
r(M)=\lambda_{\mathrm{exec}}-\lambda_{\mathrm{edit}}e(M),
\]
where \(e(M)=d(M,C)-d(G,C)\), and \(d(\cdot,\cdot)\) is the normalized token-level Levenshtein distance used in Section~\ref{sec:eval-setup}. We set \(\lambda_{\mathrm{exec}}=0.1\) and \(\lambda_{\mathrm{edit}}=1.0\). This rewards passing repairs that stay close in size to the gold patch and \rev{penalizes} edits beyond the required change: if the edit is too large ($>0.3$), it will receive a lower reward than an incorrect edit. 

\looseness=-1 We mainly evaluate
Pass@1, excess normalized Levenshtein distance, and added cognitive complexity on out-of-domain corruptions. In addition, we measure the change in performance on LiveCodeBench v6 from the base model to check whether minimal-edit training degrades broader coding ability. 

\subsection{\revb{Results}}
\paragraph{RL gives the best out-of-domain trade-off.}
\looseness=-1 Table~\ref{tab:training} separates fitting the training corruption families from generalizing to held-out ones. We observe that SFT nearly solves the in-domain split, but its Pass@1 drops from \(0.932\) to \(0.458\) out of domain. Therefore, its small edit metrics describe only the subset of repairs that still pass, so they do not indicate a usable repair policy. \rev{SFT's slightly negative out-of-domain excess distance ($-0.008$) means it changes \emph{fewer} tokens than the reference repair in 21.3\% of its correct OOD repairs (vs.\ 8.5\% across all methods). } RL is significantly better (\(p<10^{-20}\)) than SFT on out-of-domain Pass@1. rSFT, DPO, and RL generalize more reliably than SFT. Among them, DPO is marginally highest in out-of-domain Pass@1, while RL gives almost the same correctness with substantially smaller patches. \rev{We also investigated alternative metrics including excess line-diff and syntax-tree-diff metrics, both of which give the same ranking (Spearman $\rho\approx0.91$ with $E_{\text{Lev}}$). Details are in Appendix~\ref{sec:AppendixAuxMetrics}. }

\begin{table*}[t]
  \centering
  \small
  \caption{Minimal-edit training for Qwen3-4B-Instruct-2507 on in-domain and out-of-domain corruptions. Pass@1 is computed over all 400 evaluation examples; edit metrics are averaged over passing repairs. \rev{LCB: absolute LiveCodeBench v6 score (\%); parentheses give the change from the base model (32.6\%).}}
  \label{tab:training}
  \begin{tabularx}{\textwidth}{@{}l*{7}{>{\centering\arraybackslash}X}@{}}
    \toprule
    \multirow{2}{*}{Model} & \multicolumn{3}{c}{In-domain} & \multicolumn{3}{c}{Out-of-domain} & \multirow{2}{*}{\rev{LCB (\%)} $\uparrow$} \\
    \cmidrule(lr){2-4} \cmidrule(lr){5-7}
    & Pass@1 $\uparrow$ & Excess Lev.\ $\downarrow$ & Added CC $\downarrow$ & Pass@1 $\uparrow$ & Excess Lev.\ $\downarrow$ & Added CC $\downarrow$ & \\
    \midrule
    SFT & 0.932 & 0.002 & 0.000 & 0.458 & $-$0.008 & 0.006 & \rev{17.7 ($-$14.9)} \\
    rSFT & 0.782 & 0.100 & 0.435 & 0.780 & 0.107 & 0.501 & \rev{25.7 ($-$6.9)} \\
    DPO & 0.752 & 0.021 & 0.113 & 0.787 & 0.092 & 0.348 & \rev{28.0 ($-$4.6)} \\
    RL & 0.802 & 0.046 & 0.112 & 0.782 & 0.050 & 0.185 & \rev{\textbf{33.2} ($+$0.6)} \\
    \bottomrule
  \end{tabularx}
\end{table*}

\begin{figure}[t]
  \centering
  \includegraphics[width=\columnwidth]{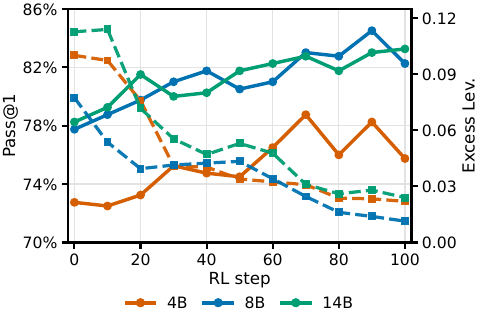}
  \caption{RL checkpoint curves for Qwen3 base models. Solid lines show Pass@1 and dashed lines show excess Levenshtein distance. Across model sizes, RL steadily reduces excess edits while preserving or improving repair success.} \vspace{-3mm}
  \label{fig:qwen3_base_rl_curves}
\end{figure}

We further scale up RL training using \rev{Qwen3} models ranging from 4B to 14B \rev{(Figure~\ref{fig:qwen3_base_rl_curves})}. 
Across all model sizes, excess Levenshtein distance falls steadily during training, while Pass@1 is preserved or improved. Thus, minimal editing can be learned as a transferable repair preference and is scalable as a learning objective. 

\paragraph{RL preserves broader coding ability.}
Minimal-edit training is useful only if it does not erase broader coding skill. Table~\ref{tab:training}'s LCB column shows that SFT loses the most \rev{(14.9 points, dropping from 32.6\% to 17.7\%)}, while rSFT and DPO also regress. RL is the only method that remains flat-to-positive \rev{(33.2\%, $+$0.6 points)}. This is consistent with recent findings that supervised fine-tuning can overfit narrow synthetic distributions, whereas reinforcement learning can better preserve transferable competence \citep{chu2025sftmemorizesrlgeneralizes,shenfeld2025rlsrazoronlinereinforcement}. In the following ablations, we only report performance on OOD setups. 

\paragraph{LoRA recovers most minimal-editing gains.}
\looseness=-1 The preceding results all use full-parameter \rev{fine-tuning}. Next, to test whether the behavior requires broad weight changes, we train Qwen3 4B with the same RL objective using LoRA adapters \citep{hu2022lora} and sweep ranks from 1 to 64. Table~\ref{tab:lora_rank} suggests that minimal editing is largely a learnable preference rather than a new coding skill. Pass@1 saturates by rank 16, but edit fidelity keeps improving through rank 64, which nearly matches full RL on excess Levenshtein distance and \rev{improves on its} added cognitive complexity. Thus, we conclude that minimal editing is a stylistic preference that can be captured with a relatively small adapter, reducing interference with the base model.

\begin{table}[t]
  \centering
  \small
  \caption{Parameter-efficient RL with LoRA. Higher ranks recover most full-parameter RL gains in edit fidelity while preserving LiveCodeBench performance.}
  \label{tab:lora_rank}
  \begin{tabularx}{\columnwidth}{@{}l*{4}{>{\centering\arraybackslash}X}@{}}
    \toprule
    Rank & Pass@1 $\uparrow$ & Excess Lev.\ $\downarrow$ & Added CC $\downarrow$ & LCB $\Delta$ $\uparrow$ \\
    \midrule
    1 & 0.738 & 0.166 & 0.676 & $-$0.022 \\
    8 & 0.775 & 0.112 & 0.426 & $-$0.022 \\
    16 & \textbf{0.805} & 0.087 & 0.328 & $-$0.005 \\
    32 & 0.795 & 0.065 & 0.235 & $-$0.011 \\
    64 & 0.797 & 0.051 & \textbf{0.160} & +0.001 \\
    Full RL & 0.782 & \textbf{0.050} & 0.185 & \textbf{+0.006} \\
    \bottomrule
  \end{tabularx}
\end{table}

\begin{table}[t]
  \centering
  \small
  \caption{Reward-design ablations for RL on Qwen3 4B Instruct. Pass@1 is computed over all examples; edit metrics are averaged over passing repairs.}
  \label{tab:reward_ablation}
  \begin{tabularx}{\columnwidth}{@{}l*{3}{>{\centering\arraybackslash}X}@{}}
    \toprule
    Reward & Pass@1 $\uparrow$ & Ex.\ Lev.\ $\downarrow$ & Add.\ CC $\downarrow$ \\
    \midrule
    Correctness only & 0.735 & 0.189 & 0.677 \\
    Edit only & 0.775 & 0.070 & 0.284 \\
    Lev.\ + CogC & 0.770 & 0.151 & 0.769 \\
    Full reward & \textbf{0.782} & \textbf{0.050} & \textbf{0.185} \\
    \bottomrule
  \end{tabularx}
\end{table}

\paragraph{Reward design trades off correctness and edit locality.}
We show the ablation of reward design in Table~\ref{tab:reward_ablation}. \revb{Correctness alone (Row 1) produces working but bulky edits, whereas the edit-only reward (Row 2) still reaches high Pass@1, which is largely due to dataset construction, since a correct solution is one with minimal edits.} Adding cognitive complexity to the reward is worse here, likely because it is too coarse for localized bug repair. The full reward gives the best balance: it keeps Pass@1 high while yielding the smallest patches and lowest added complexity.

\paragraph{\rev{Comparison with related work.}} \looseness=-1 \rev{A related work, AdaPatcher \citep{dai2025adapatcher}, pairs a runtime-trace bug locator with DPO-positive (DPOP) preference learning to favor small modifications. Lacking runtime traces, we compare against its preference-learning stage by training Qwen3-4B with DPOP under the DPO protocol. DPOP performs comparably to DPO and below RL: full-parameter DPOP reaches Pass@1 $0.718$ with excess distance $0.064$ (RL: $0.782$/$0.050$), and LoRA $r{=}64$ DPOP reaches $0.783$/$0.082$ vs.\ RL's $0.797$/$0.051$ (Appendix~\ref{sec:AppendixAdaPatcher}). This is consistent with on-policy RL generalizing better than offline preference optimization \citep{chu2025sftmemorizesrlgeneralizes}.}

\paragraph{\rev{Transfer to real bugs.}}
\looseness=-1 \rev{Finally, we evaluate the RL-trained Qwen3 models on Defects4J \citep{just2014defects4j} bugs confined to a single modified class and method. These are real human-written Java bugs, unlike our injected Python training bugs. As Table~\ref{tab:defects4j} shows, RL leaves pass rates essentially unchanged while shrinking excess Levenshtein distance and raw token edits. The minimal-editing preference therefore transfers to realistic bugs in an unseen language, though absolute repair rates remain low at these model sizes.}

\begin{table}[t]
  \centering
  \small  \caption{\rev{Cross-domain evaluation on single-method Defects4J bugs. RL preserves the pass rate while reducing edit size.}}
  \label{tab:defects4j}
  \begin{tabularx}{\columnwidth}{@{}l*{4}{>{\centering\arraybackslash}X}@{}}
    \toprule
    Model & Pass rate & Token edits $\downarrow$ & Ex.\ Lev.\ $\downarrow$ & Add.\ CC $\downarrow$ \\
    \midrule
    4B base & 7.4\% & 51.9 & 0.114 & 0.123 \\
    4B RL & 7.0\% & 35.3 & 0.060 & 0.065 \\
    14B base & 14.9\% & 40.3 & 0.105 & 0.149 \\
    14B RL & 14.5\% & 34.5 & 0.074 & 0.120 \\
    \bottomrule
  \end{tabularx}
\end{table}

\section{Discussion}

\looseness=-1 \textbf{Why edit size matters.} Software work splits roughly into \rev{greenfield} development, where new code is written from scratch, and \rev{brownfield} maintenance, where existing code must be changed without disrupting surrounding behavior. In \rev{brownfield} repair, the model's job is to fix an error, not to rewrite style or structure that was not broken. Over-editing is therefore a maintenance failure mode: unlike an incorrect answer, a gratuitously large rewrite can still pass tests, making it largely invisible to correctness-only evaluation. Reviewers must still determine what changed, whether the change is safe, and how it relates to the original implementation. A correct but unnecessarily large \rev{diff} raises review cost and \rev{hides} regressions outside the tests. We argue that the size of the edit is an important dimension \rev{alongside} functional correctness. 

\section{Conclusion} 
\looseness=-1 \rev{By measuring coding LLMs under a controlled minimal-repair ground truth, we find that over-editing is widespread.} Explicit instructions to preserve the original implementation substantially reduce excess edits for most models, with \rev{both} reasoning and open-weight coding models benefiting. Over-editing is therefore a default editing behavior, not an unavoidable capability limit. Post-training experiments further indicate that minimal editing can be learned more durably: in our study, RL improves out-of-domain edit fidelity while preserving broader coding competence better than standard supervised \rev{fine-tuning}. Overall, we believe that code-repair models should be evaluated on both functional correctness and \rev{edit fidelity}.

\section*{Limitations}
\looseness=-1 We use controlled function-level corruptions with a known minimal reversal, which yields clean ground truth for patch size but is simpler than real repository bugs and multi-file changes. \rev{The main evaluation tasks are in Python, and the training experiments use the Qwen model family. Our human studies are small in scale: three annotators over 100 patch pairs for metric validation and a single-annotator audit of 100 high-excess repairs. While the Defects4J transfer results suggest that the learned minimal-editing preference generalizes to real Java bugs, absolute repair rates in that setting remain low for the model sizes we train.} Extending the same measurements to repository-level edits, more languages, richer reward designs, and \rev{larger-scale} human evaluation is important future work.

\section*{Ethical Considerations}
\looseness=-1 Our work builds on open-source datasets and evaluation protocols. \rev{While over-editing by code agents could reduce efficiency, we do not believe this work poses significant ethical risks.} We do not think there are other ethical considerations worth mentioning.

\bibliography{main}

@misc{zhuo2025bigcodebenchbenchmarkingcodegeneration,
      title={BigCodeBench: Benchmarking Code Generation with Diverse Function Calls and Complex Instructions}, 
      author={Terry Yue Zhuo and Minh Chien Vu and Jenny Chim and Han Hu and Wenhao Yu and Ratnadira Widyasari and Imam Nur Bani Yusuf and Haolan Zhan and Junda He and Indraneil Paul and Simon Brunner and Chen Gong and Thong Hoang and Armel Randy Zebaze and Xiaoheng Hong and Wen-Ding Li and Jean Kaddour and Ming Xu and Zhihan Zhang and Prateek Yadav and Naman Jain and Alex Gu and Zhoujun Cheng and Jiawei Liu and Qian Liu and Zijian Wang and Binyuan Hui and Niklas Muennighoff and David Lo and Daniel Fried and Xiaoning Du and Harm de Vries and Leandro Von Werra},
      year={2025},
      eprint={2406.15877},
      archivePrefix={arXiv},
      primaryClass={cs.SE},
      url={https://arxiv.org/abs/2406.15877}, 
}

@article{Evtikhiev_2023,
   title={Out of the {BLEU}: How should we assess quality of the Code Generation models?},
   volume={203},
   ISSN={0164-1212},
   url={http://dx.doi.org/10.1016/j.jss.2023.111741},
   DOI={10.1016/j.jss.2023.111741},
   journal={Journal of Systems and Software},
   publisher={Elsevier BV},
   author={Evtikhiev, Mikhail and Bogomolov, Egor and Sokolov, Yaroslav and Bryksin, Timofey},
   year={2023},
   month=sep,
   pages={111741}
}

@misc{ren2020codebleumethodautomaticevaluation,
      title={CodeBLEU: a Method for Automatic Evaluation of Code Synthesis}, 
      author={Shuo Ren and Daya Guo and Shuai Lu and Long Zhou and Shujie Liu and Duyu Tang and Neel Sundaresan and Ming Zhou and Ambrosio Blanco and Shuai Ma},
      year={2020},
      eprint={2009.10297},
      archivePrefix={arXiv},
      primaryClass={cs.SE},
      url={https://arxiv.org/abs/2009.10297}, 
}

@misc{campbell2021cognitivecomplexity,
  title        = {{Cognitive Complexity}},
  author       = {Campbell, G. Ann},
  year         = {2021},
  howpublished = {\url{https://www.sonarsource.com/resources/cognitive-complexity/}},
  note         = {SonarSource white paper on the Cognitive Complexity metric.}
}

@misc{anthropic2025_claude_sonnet4,
  title        = {{Claude Sonnet 4 / Claude 4 System Card}},
  author       = {{Anthropic}},
  year         = {2025},
  howpublished = {\url{https://www.anthropic.com/claude-4-model-card}},
  note         = {Model announcement / system card.}
}

@misc{openai2025_gpt5_systemcard,
  title        = {{GPT-5 System Card}},
  author       = {{OpenAI}},
  year         = {2025},
  howpublished = {\url{https://openai.com/index/gpt-5-system-card/}},
  note         = {System card and model documentation.}
}

@misc{chen2021evaluatinglargelanguagemodels,
      title={Evaluating Large Language Models Trained on Code}, 
      author={Mark Chen and Jerry Tworek and Heewoo Jun and Qiming Yuan and Henrique Ponde de Oliveira Pinto and Jared Kaplan and Harri Edwards and Yuri Burda and Nicholas Joseph and Greg Brockman and Alex Ray and Raul Puri and Gretchen Krueger and Michael Petrov and Heidy Khlaaf and Girish Sastry and Pamela Mishkin and Brooke Chan and Scott Gray and Nick Ryder and Mikhail Pavlov and Alethea Power and Lukasz Kaiser and Mohammad Bavarian and Clemens Winter and Philippe Tillet and Felipe Petroski Such and Dave Cummings and Matthias Plappert and Fotios Chantzis and Elizabeth Barnes and Ariel Herbert-Voss and William Hebgen Guss and Alex Nichol and Alex Paino and Nikolas Tezak and Jie Tang and Igor Babuschkin and Suchir Balaji and Shantanu Jain and William Saunders and Christopher Hesse and Andrew N. Carr and Jan Leike and Josh Achiam and Vedant Misra and Evan Morikawa and Alec Radford and Matthew Knight and Miles Brundage and Mira Murati and Katie Mayer and Peter Welinder and Bob McGrew and Dario Amodei and Sam McCandlish and Ilya Sutskever and Wojciech Zaremba},
      year={2021},
      eprint={2107.03374},
      archivePrefix={arXiv},
      primaryClass={cs.LG},
      url={https://arxiv.org/abs/2107.03374}, 
}

@misc{jain2024livecodebenchholisticcontaminationfree,
      title={LiveCodeBench: Holistic and Contamination Free Evaluation of Large Language Models for Code}, 
      author={Naman Jain and King Han and Alex Gu and Wen-Ding Li and Fanjia Yan and Tianjun Zhang and Sida Wang and Armando Solar-Lezama and Koushik Sen and Ion Stoica},
      year={2024},
      eprint={2403.07974},
      archivePrefix={arXiv},
      primaryClass={cs.SE},
      url={https://arxiv.org/abs/2403.07974}, 
}

@misc{guo2025codeeditorbenchevaluatingcodeediting,
      title={CodeEditorBench: Evaluating Code Editing Capability of Large Language Models}, 
      author={Jiawei Guo and Ziming Li and Xueling Liu and Kaijing Ma and Tianyu Zheng and Zhouliang Yu and Ding Pan and Yizhi LI and Ruibo Liu and Yue Wang and Shuyue Guo and Xingwei Qu and Xiang Yue and Ge Zhang and Wenhu Chen and Jie Fu},
      year={2025},
      eprint={2404.03543},
      archivePrefix={arXiv},
      primaryClass={cs.SE},
      url={https://arxiv.org/abs/2404.03543}, 
}

@misc{cassano2024canitedit,
      title={Can It Edit? Evaluating the Ability of Large Language Models to Follow Code Editing Instructions},
      author={Federico Cassano and Luisa Li and Akul Sethi and Noah Shinn and Abby Brennan-Jones and Jacob Ginesin and Edward Berman and George Chakhnashvili and Anton Lozhkov and Carolyn Jane Anderson and Arjun Guha},
      year={2024},
      eprint={2312.12450},
      archivePrefix={arXiv},
      primaryClass={cs.SE},
      url={https://arxiv.org/abs/2312.12450},
}

@misc{chi2025editbench,
      title={{EDIT-Bench}: Evaluating LLM Abilities to Perform Real-World Instructed Code Edits},
      author={Wayne Chi and Valerie Chen and Ryan Shar and Aditya Mittal and Jenny Liang and Wei-Lin Chiang and Anastasios Nikolas Angelopoulos and Ion Stoica and Graham Neubig and Ameet Talwalkar and Chris Donahue},
      year={2025},
      eprint={2511.04486},
      archivePrefix={arXiv},
      primaryClass={cs.SE},
      url={https://arxiv.org/abs/2511.04486},
}

@misc{ebrahimi2026editverify,
      title={Edit, But Verify: An Empirical Audit of Instructed Code-Editing Benchmarks},
      author={Amir M. Ebrahimi and Gopi Krishnan Rajbahadur},
      year={2026},
      eprint={2604.05100},
      archivePrefix={arXiv},
      primaryClass={cs.SE},
      url={https://arxiv.org/abs/2604.05100},
}

@misc{yang2024cref,
      title={{CREF}: An LLM-based Conversational Software Repair Framework for Programming Tutors},
      author={Boyang Yang and Haoye Tian and Weiguo Pian and Haoran Yu and Haitao Wang and Jacques Klein and Tegawend{\'e} F. Bissyand{\'e} and Shunfu Jin},
      year={2024},
      eprint={2406.13972},
      archivePrefix={arXiv},
      primaryClass={cs.SE},
      url={https://arxiv.org/abs/2406.13972},
}

@misc{dai2025adapatcher,
      title={Less is More: Adaptive Program Repair with Bug Localization and Preference Learning},
      author={Zhenlong Dai and Bingrui Chen and Zhuoluo Zhao and Xiu Tang and Sai Wu and Chang Yao and Zhipeng Gao and Jingyuan Chen},
      year={2025},
      eprint={2503.06510},
      archivePrefix={arXiv},
      primaryClass={cs.SE},
      url={https://arxiv.org/abs/2503.06510},
}

@misc{yang2026paft,
      title={{PAFT}: Preservation Aware Fine-Tuning for Minimal-Edit Program Repair},
      author={Boyang Yang and Zijian Cai and Shunfu Jin and Haoye Tian},
      year={2026},
      eprint={2604.03113},
      archivePrefix={arXiv},
      primaryClass={cs.SE},
      url={https://arxiv.org/abs/2604.03113},
}

@misc{ke2026prepair,
      title={{QiMeng-PRepair}: Precise Code Repair via Edit-Aware Reward Optimization},
      author={Changxin Ke and Rui Zhang and Jiaming Guo and Yuanbo Wen and Li Ding and Shuo Wang and Xuyuan Zhu and Xiong Peng and Di Huang and Zidong Du and Xing Hu and Qi Guo and Yunji Chen},
      year={2026},
      eprint={2604.05963},
      archivePrefix={arXiv},
      primaryClass={cs.SE},
      url={https://arxiv.org/abs/2604.05963},
}

@misc{muennighoff2023octopack,
      title={{OctoPack}: Instruction Tuning Code Large Language Models},
      author={Niklas Muennighoff and Qian Liu and Armel Zebaze and Qinkai Zheng and Binyuan Hui and Terry Yue Zhuo and Swayam Singh and Xiangru Tang and Leandro von Werra and Shayne Longpre},
      year={2023},
      eprint={2308.07124},
      archivePrefix={arXiv},
      primaryClass={cs.CL},
      url={https://arxiv.org/abs/2308.07124},
}

@inproceedings{qi2015plausibility,
      title={An Analysis of Patch Plausibility and Correctness for Generate-and-Validate Patch Generation Systems},
      author={Zichao Qi and Fan Long and Sara Achour and Martin Rinard},
      booktitle={Proceedings of the 2015 International Symposium on Software Testing and Analysis},
      pages={24--36},
      year={2015},
      publisher={ACM},
      doi={10.1145/2771783.2771791},
      url={https://doi.org/10.1145/2771783.2771791},
}

@article{liu2021criticalreview,
      title={A Critical Review on the Evaluation of Automated Program Repair Systems},
      author={Kui Liu and Li Li and Anil Koyuncu and Dongsun Kim and Zhe Liu and Jacques Klein and Tegawend{\'e} F. Bissyand{\'e}},
      journal={Journal of Systems and Software},
      volume={171},
      pages={110817},
      year={2021},
      doi={10.1016/j.jss.2020.110817},
      url={https://doi.org/10.1016/j.jss.2020.110817},
}

@misc{jimenez2024swebenchlanguagemodelsresolve,
      title={SWE-bench: Can Language Models Resolve Real-World GitHub Issues?}, 
      author={Carlos E. Jimenez and John Yang and Alexander Wettig and Shunyu Yao and Kexin Pei and Ofir Press and Karthik Narasimhan},
      year={2024},
      eprint={2310.06770},
      archivePrefix={arXiv},
      primaryClass={cs.CL},
      url={https://arxiv.org/abs/2310.06770}, 
}

@misc{Chowdhury2024swebenchverified,
      title={Introducing SWE-Bench Verified}, 
      author={Neil Chowdhury and James Aung and Chan Jun Shern and Oliver Jaffe and Dane Sherburn and Giulio Starace and Evan Mays and Rachel Dias and Marwan Aljubeh and Mia Glaese and Carlos E. Jimenez and John Yang and Leyton Ho and Tejal Patwardhan and Kevin Liu and Aleksander Madry},
      year={2024},
      url={https://openai.com/index/introducing-swe-bench-verified/}, 
}

@misc{liu2023codegeneratedchatgptreally,
      title={Is Your Code Generated by ChatGPT Really Correct? Rigorous Evaluation of Large Language Models for Code Generation}, 
      author={Jiawei Liu and Chunqiu Steven Xia and Yuyao Wang and Lingming Zhang},
      year={2023},
      eprint={2305.01210},
      archivePrefix={arXiv},
      primaryClass={cs.SE},
      url={https://arxiv.org/abs/2305.01210}, 
}

@misc{austin2021programsynthesislargelanguage,
      title={Program Synthesis with Large Language Models}, 
      author={Jacob Austin and Augustus Odena and Maxwell Nye and Maarten Bosma and Henryk Michalewski and David Dohan and Ellen Jiang and Carrie Cai and Michael Terry and Quoc Le and Charles Sutton},
      year={2021},
      eprint={2108.07732},
      archivePrefix={arXiv},
      primaryClass={cs.PL},
      url={https://arxiv.org/abs/2108.07732}, 
}

@misc{li2023tacotopicsalgorithmiccode,
      title={TACO: Topics in Algorithmic COde generation dataset}, 
      author={Rongao Li and Jie Fu and Bo-Wen Zhang and Tao Huang and Zhihong Sun and Chen Lyu and Guang Liu and Zhi Jin and Ge Li},
      year={2023},
      eprint={2312.14852},
      archivePrefix={arXiv},
      primaryClass={cs.AI},
      url={https://arxiv.org/abs/2312.14852}, 
}

@misc{ni2023l2cevalevaluatinglanguagetocodegeneration,
      title={L2CEval: Evaluating Language-to-Code Generation Capabilities of Large Language Models}, 
      author={Ansong Ni and Pengcheng Yin and Yilun Zhao and Martin Riddell and Troy Feng and Rui Shen and Stephen Yin and Ye Liu and Semih Yavuz and Caiming Xiong and Shafiq Joty and Yingbo Zhou and Dragomir Radev and Arman Cohan},
      year={2023},
      eprint={2309.17446},
      archivePrefix={arXiv},
      primaryClass={cs.CL},
      url={https://arxiv.org/abs/2309.17446}, 
}

@article{Zheng2024HumanEvoAEB,
  title={HumanEvo: An Evolution-Aware Benchmark for More Realistic Evaluation of Repository-Level Code Generation},
  author={Dewu Zheng and Yanlin Wang and Ensheng Shi and Ruikai Zhang and Yuchi Ma and Hongyu Zhang and Zibin Zheng},
  journal={2025 IEEE/ACM 47th International Conference on Software Engineering (ICSE)},
  year={2024},
  pages={1372-1384},
  eprint={2406.06918},
  archivePrefix={arXiv},
  primaryClass={cs.SE},
  url={https://arxiv.org/abs/2406.06918}
}

@article{He2025SWEPerfCLD,
  title={SWE-Perf: Can Language Models Optimize Code Performance on Real-World Repositories?},
  author={Xinyi He and Qian Liu and Mingzhe Du and Lin Yan and Zhijie Fan and Yiming Huang and Zejian Yuan and Zejun Ma},
  journal={ArXiv},
  year={2025},
  volume={abs/2507.12415},
  eprint={2507.12415},
  archivePrefix={arXiv},
  primaryClass={cs.SE},
  url={https://arxiv.org/abs/2507.12415}
}

@misc{fu2025corecodebenchconfigurablemultiscenariorepositorylevel,
      title={CoreCodeBench: Decoupling Code Intelligence via Fine-Grained Repository-Level Tasks},
      author={Lingyue Fu and Hao Guan and Bolun Zhang and Haowei Yuan and Yaoming Zhu and Jun Xu and Zongyu Wang and Lin Qiu and Xunliang Cai and Xuezhi Cao and Weiwen Liu and Weinan Zhang and Yong Yu},
      year={2025},
      eprint={2507.05281},
      archivePrefix={arXiv},
      primaryClass={cs.SE},
      url={https://arxiv.org/abs/2507.05281}, 
}

@misc{tian2024debugbenchevaluatingdebuggingcapability,
      title={DebugBench: Evaluating Debugging Capability of Large Language Models}, 
      author={Runchu Tian and Yining Ye and Yujia Qin and Xin Cong and Yankai Lin and Yinxu Pan and Yesai Wu and Haotian Hui and Weichuan Liu and Zhiyuan Liu and Maosong Sun},
      year={2024},
      eprint={2401.04621},
      archivePrefix={arXiv},
      primaryClass={cs.SE},
      url={https://arxiv.org/abs/2401.04621}, 
}

@misc{gauthier2024aider,
  author = {Gauthier, Paul},
  title = {GPT code editing benchmarks},
  howpublished = {\url{https://aider.chat/docs/benchmarks.html}},
  year = {2024},
  note = {Accessed: 2024}
}

@misc{zhou2023instructionfollowingevaluationlargelanguage,
      title={Instruction-Following Evaluation for Large Language Models}, 
      author={Jeffrey Zhou and Tianjian Lu and Swaroop Mishra and Siddhartha Brahma and Sujoy Basu and Yi Luan and Denny Zhou and Le Hou},
      year={2023},
      eprint={2311.07911},
      archivePrefix={arXiv},
      primaryClass={cs.CL},
      url={https://arxiv.org/abs/2311.07911}, 
}

@misc{fu2025scalingreasoninglosingcontrol,
      title={Scaling Reasoning, Losing Control: Evaluating Instruction Following in Large Reasoning Models}, 
      author={Tingchen Fu and Jiawei Gu and Yafu Li and Xiaoye Qu and Yu Cheng},
      year={2025},
      eprint={2505.14810},
      archivePrefix={arXiv},
      primaryClass={cs.CL},
      url={https://arxiv.org/abs/2505.14810}, 
}

@misc{wen2024benchmarkingcomplexinstructionfollowingmultiple,
      title={Benchmarking Complex Instruction-Following with Multiple Constraints Composition}, 
      author={Bosi Wen and Pei Ke and Xiaotao Gu and Lindong Wu and Hao Huang and Jinfeng Zhou and Wenchuang Li and Binxin Hu and Wendy Gao and Jiaxin Xu and Yiming Liu and Jie Tang and Hongning Wang and Minlie Huang},
      year={2024},
      eprint={2407.03978},
      archivePrefix={arXiv},
      primaryClass={cs.CL},
      url={https://arxiv.org/abs/2407.03978}, 
}

@misc{li2025thinkingfailspitfallsreasoning,
      title={When Thinking Fails: The Pitfalls of Reasoning for Instruction-Following in LLMs}, 
      author={Xiaomin Li and Zhou Yu and Zhiwei Zhang and Xupeng Chen and Ziji Zhang and Yingying Zhuang and Narayanan Sadagopan and Anurag Beniwal},
      year={2025},
      eprint={2505.11423},
      archivePrefix={arXiv},
      primaryClass={cs.CL},
      url={https://arxiv.org/abs/2505.11423}, 
}

@misc{luo2025deepcoder,
  title        = {{DeepCoder: A Fully Open-Source 14B Coder at O3-mini Level}},
  author       = {Luo, Michael and Tan, Sijun and Huang, Roy and Patel, Ameen and Ariyak, Alpay and Wu, Qingyang and Shi, Xiaoxiang and Xin, Rachel and Cai, Colin and Weber, Maurice and Zhang, Ce and Li, Li Erran and Popa, Raluca Ada and Stoica, Ion},
  year         = {2025},
  month        = {April},
  howpublished = {Together AI Blog},
  url          = {https://www.together.ai/blog/deepcoder},
  note         = {Accessed: 2026-03-23}
}

@misc{rafailov2024directpreferenceoptimizationlanguage,
      title={Direct Preference Optimization: Your Language Model is Secretly a Reward Model}, 
      author={Rafael Rafailov and Archit Sharma and Eric Mitchell and Stefano Ermon and Christopher D. Manning and Chelsea Finn},
      year={2024},
      eprint={2305.18290},
      archivePrefix={arXiv},
      primaryClass={cs.LG},
      url={https://arxiv.org/abs/2305.18290}, 
}

@inproceedings{zheng2024llamafactory,
  title={LlamaFactory: Unified Efficient Fine-Tuning of 100+ Language Models},
  author={Yaowei Zheng and Richong Zhang and Junhao Zhang and Yanhan Ye and Zheyan Luo and Zhangchi Feng and Yongqiang Ma},
  booktitle={Proceedings of the 62nd Annual Meeting of the Association for Computational Linguistics (Volume 3: System Demonstrations)},
  address={Bangkok, Thailand},
  publisher={Association for Computational Linguistics},
  year={2024},
  url={http://arxiv.org/abs/2403.13372}
}

@misc{primeintellect2025prime-rl,
  author = {{Prime Intellect}},
  title = {{PRIME-RL}},
  url = {https://github.com/PrimeIntellect-ai/prime-rl},
  year = {2025}
}

@inproceedings{hu2022lora,
  title = {{LoRA}: Low-Rank Adaptation of Large Language Models},
  author = {Hu, Edward J. and Shen, Yelong and Wallis, Phillip and Allen-Zhu, Zeyuan and Li, Yuanzhi and Wang, Shean and Wang, Lu and Chen, Weizhu},
  booktitle = {International Conference on Learning Representations},
  year = {2022},
  url = {https://openreview.net/forum?id=nZeVKeeFYf9}
}

@misc{chu2025sftmemorizesrlgeneralizes,
      title={SFT Memorizes, RL Generalizes: A Comparative Study of Foundation Model Post-training}, 
      author={Tianzhe Chu and Yuexiang Zhai and Jihan Yang and Shengbang Tong and Saining Xie and Dale Schuurmans and Quoc V. Le and Sergey Levine and Yi Ma},
      year={2025},
      eprint={2501.17161},
      archivePrefix={arXiv},
      primaryClass={cs.AI},
      url={https://arxiv.org/abs/2501.17161}, 
}

@misc{shenfeld2025rlsrazoronlinereinforcement,
      title={RL's Razor: Why Online Reinforcement Learning Forgets Less}, 
      author={Idan Shenfeld and Jyothish Pari and Pulkit Agrawal},
      year={2025},
      eprint={2509.04259},
      archivePrefix={arXiv},
      primaryClass={cs.LG},
      url={https://arxiv.org/abs/2509.04259}, 
}

@misc{openai2026_gpt54_thinking_systemcard,
  title        = {{GPT-5.4 Thinking System Card}},
  author       = {{OpenAI}},
  year         = {2026},
  howpublished = {\url{https://openai.com/index/gpt-5-4-thinking-system-card/}},
  note         = {Official system card landing page (links to full card on OpenAI Deployment Safety Hub).}
}

@misc{openai2026_gpt55,
  title        = {{Introducing GPT-5.5}},
  author       = {{OpenAI}},
  year         = {2026},
  howpublished = {\url{https://openai.com/index/introducing-gpt-5-5/}},
  note         = {Official OpenAI model announcement.}
}

@misc{google2026_gemini31_pro,
  title        = {{Gemini 3.1 Pro: A smarter model for your most complex tasks}},
  author       = {{Google}},
  year         = {2026},
  howpublished = {\url{https://blog.google/innovation-and-ai/models-and-research/gemini-models/gemini-3-1-pro/}},
  note         = {Official Google blog announcement.}
}

@misc{xai2026_grok43,
  title        = {{Grok 4.3}},
  author       = {{xAI}},
  year         = {2026},
  howpublished = {\url{https://docs.x.ai/developers/models/grok-4.3}},
  note         = {Official xAI model documentation.}
}

@misc{zai2026_glm5,
  title        = {{GLM-5: From Vibe Coding to Agentic Engineering}},
  author       = {{Z.ai}},
  year         = {2026},
  howpublished = {\url{https://z.ai/blog/glm-5}},
  note         = {Official Z.ai technical blog post.}
}

@misc{zai2026_glm51,
  title        = {{GLM-5.1}},
  author       = {{Z.ai}},
  year         = {2026},
  howpublished = {\url{https://docs.z.ai/guides/llm/glm-5.1}},
  note         = {Official Z.ai developer documentation.}
}

@misc{alibaba2026_qwen36plus,
  title        = {{Qwen3.6-Plus: Towards Real World Agents}},
  author       = {{Alibaba Cloud}},
  year         = {2026},
  howpublished = {\url{https://www.alibabacloud.com/blog/qwen3-6-plus-towards-real-world-agents_603005}},
  note         = {Alibaba Cloud Community announcement.}
}

@misc{moonshot2026_kimi_k25,
  title        = {{Kimi K2.5 Tech Blog: Visual Agentic Intelligence}},
  author       = {{Moonshot AI}},
  year         = {2026},
  howpublished = {\url{https://www.kimi.com/blog/kimi-k2-5}},
  note         = {Official Moonshot AI / Kimi blog post.}
}

@misc{moonshot2026_kimi_k26,
  title        = {{Kimi K2.6}},
  author       = {{Moonshot AI}},
  year         = {2026},
  howpublished = {\url{https://www.kimi.com/blog/kimi-k2-6}},
  note         = {Official Moonshot AI / Kimi blog post.}
}

@misc{anthropic2026claudeopus,
  author       = {{Anthropic}},
  title        = {{Claude Opus 4.7}},
  year         = {2026},
  month        = apr,
  howpublished = {\url{https://www.anthropic.com/news/claude-opus-4-7}},
  note         = {Official Anthropic model announcement.}
}

@misc{anthropic2026claudeopus46,
  author       = {{Anthropic}},
  title        = {{Claude Opus 4.6}},
  year         = {2026},
  month        = feb,
  howpublished = {\url{https://www.anthropic.com/news/claude-opus-4-6}},
  note         = {Official Anthropic model announcement.}
}

@misc{anthropic2026_claude_sonnet46,
  title        = {{Claude Sonnet 4.6}},
  author       = {{Anthropic}},
  year         = {2026},
  howpublished = {\url{https://www.anthropic.com/news/claude-sonnet-4-6}},
  note         = {Official Anthropic model announcement.}
}

@misc{anthropic2025_claude37_sonnet,
  title        = {Claude 3.7 {S}onnet},
  author       = {{Anthropic}},
  year         = {2025},
  howpublished = {\url{https://www.anthropic.com/news/claude-3-7-sonnet}},
  note         = {Model announcement and system card.}
}

@misc{claude_sonnet_35_system_card,
  title        = {{Claude 3.5 System Card}},
  author       = {{Anthropic}},
  year         = {2024},
  howpublished = {\url{https://www.anthropic.com/news/claude-3-5-sonnet}},
  note         = {Model announcement / system card.}
}

@misc{openai2025_gpt41,
  title        = {{Introducing GPT-4.1}},
  author       = {{OpenAI}},
  year         = {2025},
  howpublished = {\url{https://openai.com/index/gpt-4-1/}},
  note         = {GPT 4.1 announcement}
}

@misc{openai2025_o4_mini,
  title        = {{Introducing o3 and o4-mini}},
  author       = {{OpenAI}},
  year         = {2025},
  howpublished = {\url{https://openai.com/index/introducing-o3-and-o4-mini/}},
  note         = {o4-mini model announcement and system card.}
}

@misc{google2025_gemini25_flash,
  title        = {{Gemini 2.5 Flash is now in preview}},
  author       = {{Google}},
  year         = {2025},
  howpublished = {\url{https://blog.google/products-and-platforms/products/gemini/gemini-2-5-flash-preview/}},
  note         = {Official Google announcement for Gemini 2.5 Flash.}
}

@misc{5team2025glm45agenticreasoningcoding,
      title={GLM-4.5: Agentic, Reasoning, and Coding (ARC) Foundation Models}, 
      author={{GLM-4.5 Team} and Aohan Zeng and Xin Lv and Qinkai Zheng and Zhenyu Hou and Bin Chen and Chengxing Xie and Cunxiang Wang and Da Yin and Hao Zeng and Jiajie Zhang and Kedong Wang and Lucen Zhong and Mingdao Liu and Rui Lu and Shulin Cao and Xiaohan Zhang and Xuancheng Huang and Yao Wei and Yean Cheng and Yifan An and Yilin Niu and Yuanhao Wen and Yushi Bai and Zhengxiao Du and Zihan Wang and Zilin Zhu and Bohan Zhang and Bosi Wen and Bowen Wu and Bowen Xu and Can Huang and Casey Zhao and Changpeng Cai and Chao Yu and Chen Li and Chendi Ge and Chenghua Huang and Chenhui Zhang and Chenxi Xu and Chenzheng Zhu and Chuang Li and Congfeng Yin and Daoyan Lin and Dayong Yang and Dazhi Jiang and Ding Ai and Erle Zhu and Fei Wang and Gengzheng Pan and Guo Wang and Hailong Sun and Haitao Li and Haiyang Li and Haiyi Hu and Hanyu Zhang and Hao Peng and Hao Tai and Haoke Zhang and Haoran Wang and Haoyu Yang and He Liu and He Zhao and Hongwei Liu and Hongxi Yan and Huan Liu and Huilong Chen and Ji Li and Jiajing Zhao and Jiamin Ren and Jian Jiao and Jiani Zhao and Jianyang Yan and Jiaqi Wang and Jiayi Gui and Jiayue Zhao and Jie Liu and Jijie Li and Jing Li and Jing Lu and Jingsen Wang and Jingwei Yuan and Jingxuan Li and Jingzhao Du and Jinhua Du and Jinxin Liu and Junkai Zhi and Junli Gao and Ke Wang and Lekang Yang and Liang Xu and Lin Fan and Lindong Wu and Lintao Ding and Lu Wang and Man Zhang and Minghao Li and Minghuan Xu and Mingming Zhao and Mingshu Zhai and Pengfan Du and Qian Dong and Shangde Lei and Shangqing Tu and Shangtong Yang and Shaoyou Lu and Shijie Li and Shuang Li and Shuang-Li and Shuxun Yang and Sibo Yi and Tianshu Yu and Wei Tian and Weihan Wang and Wenbo Yu and Weng Lam Tam and Wenjie Liang and Wentao Liu and Xiao Wang and Xiaohan Jia and Xiaotao Gu and Xiaoying Ling and Xin Wang and Xing Fan and Xingru Pan and Xinyuan Zhang and Xinze Zhang and Xiuqing Fu and Xunkai Zhang and Yabo Xu and Yandong Wu and Yida Lu and Yidong Wang and Yilin Zhou and Yiming Pan and Ying Zhang and Yingli Wang and Yingru Li and Yinpei Su and Yipeng Geng and Yitong Zhu and Yongkun Yang and Yuhang Li and Yuhao Wu and Yujiang Li and Yunan Liu and Yunqing Wang and Yuntao Li and Yuxuan Zhang and Zezhen Liu and Zhen Yang and Zhengda Zhou and Zhongpei Qiao and Zhuoer Feng and Zhuorui Liu and Zichen Zhang and Zihan Wang and Zijun Yao and Zikang Wang and Ziqiang Liu and Ziwei Chai and Zixuan Li and Zuodong Zhao and Wenguang Chen and Jidong Zhai and Bin Xu and Minlie Huang and Hongning Wang and Juanzi Li and Yuxiao Dong and Jie Tang},
      year={2025},
      eprint={2508.06471},
      archivePrefix={arXiv},
      primaryClass={cs.CL},
      url={https://arxiv.org/abs/2508.06471}, 
}

@misc{mistral2025_magistral,
  title        = {{Magistral (Mistral)---model announcement}},
  author       = {{Mistral AI}},
  year         = {2025},
  howpublished = {\url{https://mistral.ai/news/magistral}},
  note         = {Model announcement and details for Magistral (Magistral Medium).}
}

@misc{mistralai_mistral_medium3,
  title        = {{Mistral Medium 3 announcement}},
  author       = {{Mistral AI}},
  year         = {2025},
  howpublished = {\url{https://mistral.ai/news/mistral-medium-3}},
  note         = {Mistral Medium model announcement (Medium 3 family).}
}

@misc{qwen2025qwen3coder,
  title        = {{Qwen3-Coder: Agentic Coding in the World}},
  author       = {{Qwen Team}},
  year         = {2025},
  howpublished = {\url{https://qwenlm.github.io/blog/qwen3-coder/}},
  note         = {Official Qwen blog announcement.}
}

@misc{qwen2025qwen3235bthinking2507,
  title        = {{Qwen3-235B-A22B-Thinking-2507}},
  author       = {{Qwen Team}},
  year         = {2025},
  howpublished = {\url{https://huggingface.co/Qwen/Qwen3-235B-A22B-Thinking-2507}},
  note         = {Official Qwen model card.}
}

@misc{yang2025qwen3technicalreport,
      title={Qwen3 Technical Report}, 
      author={An Yang and Anfeng Li and Baosong Yang and Beichen Zhang and Binyuan Hui and Bo Zheng and Bowen Yu and Chang Gao and Chengen Huang and Chenxu Lv and Chujie Zheng and Dayiheng Liu and Fan Zhou and Fei Huang and Feng Hu and Hao Ge and Haoran Wei and Huan Lin and Jialong Tang and Jian Yang and Jianhong Tu and Jianwei Zhang and Jianxin Yang and Jiaxi Yang and Jing Zhou and Jingren Zhou and Junyang Lin and Kai Dang and Keqin Bao and Kexin Yang and Le Yu and Lianghao Deng and Mei Li and Mingfeng Xue and Mingze Li and Pei Zhang and Peng Wang and Qin Zhu and Rui Men and Ruize Gao and Shixuan Liu and Shuang Luo and Tianhao Li and Tianyi Tang and Wenbiao Yin and Xingzhang Ren and Xinyu Wang and Xinyu Zhang and Xuancheng Ren and Yang Fan and Yang Su and Yichang Zhang and Yinger Zhang and Yu Wan and Yuqiong Liu and Zekun Wang and Zeyu Cui and Zhenru Zhang and Zhipeng Zhou and Zihan Qiu},
      year={2025},
      eprint={2505.09388},
      archivePrefix={arXiv},
      primaryClass={cs.CL},
      url={https://arxiv.org/abs/2505.09388}, 
}

@misc{deepseekai2025deepseekr1incentivizingreasoningcapability,
      title={DeepSeek-R1: Incentivizing Reasoning Capability in LLMs via Reinforcement Learning}, 
      author={DeepSeek-AI and Daya Guo and Dejian Yang and Haowei Zhang and Junxiao Song and Ruoyu Zhang and Runxin Xu and Qihao Zhu and Shirong Ma and Peiyi Wang and Xiao Bi and Xiaokang Zhang and Xingkai Yu and Yu Wu and Z. F. Wu and Zhibin Gou and Zhihong Shao and Zhuoshu Li and Ziyi Gao and Aixin Liu and Bing Xue and Bingxuan Wang and Bochao Wu and Bei Feng and Chengda Lu and Chenggang Zhao and Chengqi Deng and Chenyu Zhang and Chong Ruan and Damai Dai and Deli Chen and Dongjie Ji and Erhang Li and Fangyun Lin and Fucong Dai and Fuli Luo and Guangbo Hao and Guanting Chen and Guowei Li and H. Zhang and Han Bao and Hanwei Xu and Haocheng Wang and Honghui Ding and Huajian Xin and Huazuo Gao and Hui Qu and Hui Li and Jianzhong Guo and Jiashi Li and Jiawei Wang and Jingchang Chen and Jingyang Yuan and Junjie Qiu and Junlong Li and J. L. Cai and Jiaqi Ni and Jian Liang and Jin Chen and Kai Dong and Kai Hu and Kaige Gao and Kang Guan and Kexin Huang and Kuai Yu and Lean Wang and Lecong Zhang and Liang Zhao and Litong Wang and Liyue Zhang and Lei Xu and Leyi Xia and Mingchuan Zhang and Minghua Zhang and Minghui Tang and Meng Li and Miaojun Wang and Mingming Li and Ning Tian and Panpan Huang and Peng Zhang and Qiancheng Wang and Qinyu Chen and Qiushi Du and Ruiqi Ge and Ruisong Zhang and Ruizhe Pan and Runji Wang and R. J. Chen and R. L. Jin and Ruyi Chen and Shanghao Lu and Shangyan Zhou and Shanhuang Chen and Shengfeng Ye and Shiyu Wang and Shuiping Yu and Shunfeng Zhou and Shuting Pan and S. S. Li and Shuang Zhou and Shaoqing Wu and Shengfeng Ye and Tao Yun and Tian Pei and Tianyu Sun and T. Wang and Wangding Zeng and Wanjia Zhao and Wen Liu and Wenfeng Liang and Wenjun Gao and Wenqin Yu and Wentao Zhang and W. L. Xiao and Wei An and Xiaodong Liu and Xiaohan Wang and Xiaokang Chen and Xiaotao Nie and Xin Cheng and Xin Liu and Xin Xie and Xingchao Liu and Xinyu Yang and Xinyuan Li and Xuecheng Su and Xuheng Lin and X. Q. Li and Xiangyue Jin and Xiaojin Shen and Xiaosha Chen and Xiaowen Sun and Xiaoxiang Wang and Xinnan Song and Xinyi Zhou and Xianzu Wang and Xinxia Shan and Y. K. Li and Y. Q. Wang and Y. X. Wei and Yang Zhang and Yanhong Xu and Yao Li and Yao Zhao and Yaofeng Sun and Yaohui Wang and Yi Yu and Yichao Zhang and Yifan Shi and Yiliang Xiong and Ying He and Yishi Piao and Yisong Wang and Yixuan Tan and Yiyang Ma and Yiyuan Liu and Yongqiang Guo and Yuan Ou and Yuduan Wang and Yue Gong and Yuheng Zou and Yujia He and Yunfan Xiong and Yuxiang Luo and Yuxiang You and Yuxuan Liu and Yuyang Zhou and Y. X. Zhu and Yanhong Xu and Yanping Huang and Yaohui Li and Yi Zheng and Yuchen Zhu and Yunxian Ma and Ying Tang and Yukun Zha and Yuting Yan and Z. Z. Ren and Zehui Ren and Zhangli Sha and Zhe Fu and Zhean Xu and Zhenda Xie and Zhengyan Zhang and Zhewen Hao and Zhicheng Ma and Zhigang Yan and Zhiyu Wu and Zihui Gu and Zijia Zhu and Zijun Liu and Zilin Li and Ziwei Xie and Ziyang Song and Zizheng Pan and Zhen Huang and Zhipeng Xu and Zhongyu Zhang and Zhen Zhang},
      year={2025},
      eprint={2501.12948},
      archivePrefix={arXiv},
      primaryClass={cs.CL},
      url={https://arxiv.org/abs/2501.12948}, 
}

@misc{deepseekai2025deepseekv32,
      title={DeepSeek-V3.2: Pushing the Frontier of Open Large Language Models},
      author={DeepSeek-AI},
      year={2025},
      eprint={2512.02556},
      archivePrefix={arXiv},
      primaryClass={cs.CL},
      url={https://arxiv.org/abs/2512.02556},
}

@misc{deepseekai2025deepseekv3technicalreport,
      title={DeepSeek-V3 Technical Report}, 
      author={DeepSeek-AI and Aixin Liu and Bei Feng and Bing Xue and Bingxuan Wang and Bochao Wu and Chengda Lu and Chenggang Zhao and Chengqi Deng and Chenyu Zhang and Chong Ruan and Damai Dai and Daya Guo and Dejian Yang and Deli Chen and Dongjie Ji and Erhang Li and Fangyun Lin and Fucong Dai and Fuli Luo and Guangbo Hao and Guanting Chen and Guowei Li and H. Zhang and Han Bao and Hanwei Xu and Haocheng Wang and Haowei Zhang and Honghui Ding and Huajian Xin and Huazuo Gao and Hui Li and Hui Qu and J. L. Cai and Jian Liang and Jianzhong Guo and Jiaqi Ni and Jiashi Li and Jiawei Wang and Jin Chen and Jingchang Chen and Jingyang Yuan and Junjie Qiu and Junlong Li and Junxiao Song and Kai Dong and Kai Hu and Kaige Gao and Kang Guan and Kexin Huang and Kuai Yu and Lean Wang and Lecong Zhang and Lei Xu and Leyi Xia and Liang Zhao and Litong Wang and Liyue Zhang and Meng Li and Miaojun Wang and Mingchuan Zhang and Minghua Zhang and Minghui Tang and Mingming Li and Ning Tian and Panpan Huang and Peiyi Wang and Peng Zhang and Qiancheng Wang and Qihao Zhu and Qinyu Chen and Qiushi Du and R. J. Chen and R. L. Jin and Ruiqi Ge and Ruisong Zhang and Ruizhe Pan and Runji Wang and Runxin Xu and Ruoyu Zhang and Ruyi Chen and S. S. Li and Shanghao Lu and Shangyan Zhou and Shanhuang Chen and Shaoqing Wu and Shengfeng Ye and Shengfeng Ye and Shirong Ma and Shiyu Wang and Shuang Zhou and Shuiping Yu and Shunfeng Zhou and Shuting Pan and T. Wang and Tao Yun and Tian Pei and Tianyu Sun and W. L. Xiao and Wangding Zeng and Wanjia Zhao and Wei An and Wen Liu and Wenfeng Liang and Wenjun Gao and Wenqin Yu and Wentao Zhang and X. Q. Li and Xiangyue Jin and Xianzu Wang and Xiao Bi and Xiaodong Liu and Xiaohan Wang and Xiaojin Shen and Xiaokang Chen and Xiaokang Zhang and Xiaosha Chen and Xiaotao Nie and Xiaowen Sun and Xiaoxiang Wang and Xin Cheng and Xin Liu and Xin Xie and Xingchao Liu and Xingkai Yu and Xinnan Song and Xinxia Shan and Xinyi Zhou and Xinyu Yang and Xinyuan Li and Xuecheng Su and Xuheng Lin and Y. K. Li and Y. Q. Wang and Y. X. Wei and Y. X. Zhu and Yang Zhang and Yanhong Xu and Yanhong Xu and Yanping Huang and Yao Li and Yao Zhao and Yaofeng Sun and Yaohui Li and Yaohui Wang and Yi Yu and Yi Zheng and Yichao Zhang and Yifan Shi and Yiliang Xiong and Ying He and Ying Tang and Yishi Piao and Yisong Wang and Yixuan Tan and Yiyang Ma and Yiyuan Liu and Yongqiang Guo and Yu Wu and Yuan Ou and Yuchen Zhu and Yuduan Wang and Yue Gong and Yuheng Zou and Yujia He and Yukun Zha and Yunfan Xiong and Yunxian Ma and Yuting Yan and Yuxiang Luo and Yuxiang You and Yuxuan Liu and Yuyang Zhou and Z. F. Wu and Z. Z. Ren and Zehui Ren and Zhangli Sha and Zhe Fu and Zhean Xu and Zhen Huang and Zhen Zhang and Zhenda Xie and Zhengyan Zhang and Zhewen Hao and Zhibin Gou and Zhicheng Ma and Zhigang Yan and Zhihong Shao and Zhipeng Xu and Zhiyu Wu and Zhongyu Zhang and Zhuoshu Li and Zihui Gu and Zijia Zhu and Zijun Liu and Zilin Li and Ziwei Xie and Ziyang Song and Ziyi Gao and Zizheng Pan},
      year={2025},
      eprint={2412.19437},
      archivePrefix={arXiv},
      primaryClass={cs.CL},
      url={https://arxiv.org/abs/2412.19437}, 
}

@misc{deepseek_v31_2025,
  title        = {{DeepSeek-V3.1 Release}},
  author       = {{DeepSeek AI}},
  year         = {2025},
  howpublished = {\url{https://api-docs.deepseek.com/news/news250821}},
  note         = {Model listing and documentation.}
}

@misc{hui2024qwen25coder,
      title={Qwen2.5-Coder Technical Report},
      author={Binyuan Hui and Jian Yang and Zeyu Cui and Jiaxi Yang and Dayiheng Liu and Lei Zhang and Tianyu Liu and Jiajun Zhang and Bowen Yu and Keming Lu and Kai Dang and Yang Fan and Yichang Zhang and An Yang and Rui Men and Fei Huang and Bo Zheng and Yibo Miao and Shanghaoran Quan and Yunlong Feng and Xingzhang Ren and Xuancheng Ren and Jingren Zhou and Junyang Lin},
      year={2024},
      eprint={2409.12186},
      archivePrefix={arXiv},
      primaryClass={cs.CL},
      url={https://arxiv.org/abs/2409.12186},
}

@misc{meta2024llama31,
  title        = {{Meta Llama 3.1 70B Instruct}},
  author       = {{Meta AI}},
  year         = {2024},
  howpublished = {\url{https://huggingface.co/meta-llama/Llama-3.1-70B-Instruct}},
  note         = {Official model card and license information.}
}

@inproceedings{just2014defects4j,
author = {Just, Ren{\'e} and Jalali, Darioush and Ernst, Michael D.},
title = {Defects4J: a database of existing faults to enable controlled testing studies for Java programs},
year = {2014},
isbn = {9781450326452},
publisher = {Association for Computing Machinery},
address = {New York, NY, USA},
url = {https://doi.org/10.1145/2610384.2628055},
doi = {10.1145/2610384.2628055},
booktitle = {Proceedings of the 2014 International Symposium on Software Testing and Analysis},
pages = {437–440},
numpages = {4},
location = {San Jose, CA, USA},
series = {ISSTA 2014}
}

@misc{chai2026precise,
  title = {Precise Debugging Benchmark: Is Your Model Debugging or Regenerating?},
  author = {Miaosen Chai and Wang Bill Zhu and Shangshang Wang and Yejia Liu and Song Bian and Honghua Dong and Willie Neiswanger and Robin Jia},
  year = {2026},
  eprint = {2604.17338},
  archivePrefix = {arXiv},
  primaryClass = {cs.SE},
  url = {https://arxiv.org/abs/2604.17338}
}

\appendix
\renewcommand{\topfraction}{0.95}
\renewcommand{\bottomfraction}{0.95}
\renewcommand{\textfraction}{0.03}
\renewcommand{\floatpagefraction}{0.75}
\setcounter{topnumber}{4}
\setcounter{bottomnumber}{4}
\setcounter{totalnumber}{8}
\setlength{\floatsep}{8pt plus 2pt minus 2pt}
\setlength{\textfloatsep}{8pt plus 2pt minus 2pt}
\setlength{\dblfloatsep}{8pt plus 2pt minus 2pt}
\setlength{\dbltextfloatsep}{8pt plus 2pt minus 2pt}
\setlength{\abovecaptionskip}{4pt}
\setlength{\belowcaptionskip}{0pt}

\section*{Appendix}
\rev{\noindent The appendix has five parts. Appendix~\ref{sec:AppendixBenchmarkProtocol} documents how we construct the repair benchmark and query models. Appendix~\ref{sec:AppendixValidation} reports the checks that validate our edit-fidelity measurements, including two human studies. Appendix~\ref{sec:AppendixEvalResults} gives the per-model evaluation results and the statistical robustness analyses behind the preservation-prompt effect. Appendix~\ref{sec:AppendixTraining} covers the post-training experiments and their ablations. Appendix~\ref{sec:AppendixResponsibleResearch} includes the responsible-research statements.}


\section{Benchmark Construction and Evaluation Protocol}
\label{sec:AppendixBenchmarkProtocol}
\rev{This section documents how we build the repair benchmark and how we query models: the corruption families we inject into the reference solutions, the composition of the resulting 400-task evaluation set, the exact prompts for the two conditions, and the API and decoding settings.}

\subsection{BigCodeBench Corruption Families}
\label{sec:AppendixCorruptions}
For each evaluation example, we sample \rev{one or two} corruptions from the families below.
\begin{enumerate}
    \setlength{\itemsep}{0pt}
    \setlength{\parsep}{0pt}
    \item \textbf{Comparison Operators} -- Replace relational operators (e.g., \texttt{<} $\leftrightarrow$ \texttt{<=}, \texttt{==} $\leftrightarrow$ \texttt{!=}, \texttt{>} $\leftrightarrow$ \texttt{>=}).
    \item \looseness=-1 \textbf{Range Bounds} -- Add off-by-one errors in \texttt{range()} calls (\texttt{range(n)} $\leftrightarrow$ \texttt{range(n+1)}).
    \item \textbf{Sort Order} -- Invert sorting order by toggling \texttt{reverse=True} $\leftrightarrow$ \texttt{reverse=False}.
    \item \looseness=-1 \textbf{Accumulator Initialization} -- Modify initial values of accumulators (\texttt{0} $\rightarrow$ \texttt{1}, \texttt{[]} $\rightarrow$ \texttt{[0]}).
    \item \looseness=-1 \textbf{Arithmetic Operators} -- Substitute arithmetic operators (e.g., \texttt{+} $\leftrightarrow$ \texttt{-}, \texttt{*} $\leftrightarrow$ \texttt{//}, \texttt{/} $\leftrightarrow$ \texttt{*}).
    \item \textbf{Edge Case Guards} -- Add, modify, or remove edge-case handling conditions.
    \item \textbf{List Indexing} -- Introduce off-by-one errors in list or array indexing.
    \item \textbf{Function Call Substitution} -- Replace function calls with semantically similar ones (e.g., \texttt{mean} $\leftrightarrow$ \texttt{median}, \texttt{max} $\leftrightarrow$ \texttt{min}, \texttt{sum} $\leftrightarrow$ \texttt{len}).
    \item \textbf{Copy Removal} -- Remove calls to \texttt{.copy()} methods, altering reference semantics.
    \item \textbf{Boolean Constants} -- Flip boolean constants (\texttt{True} $\leftrightarrow$ \texttt{False}).
    \item \looseness=-1 \textbf{Numeric Constants} -- Adjust numeric constants (integers by $\pm1$, floats by $\pm0.1$).
    \item \textbf{Slice Bounds} -- Shift slice boundaries by one (e.g., \texttt{[a:b]} $\leftrightarrow$ \texttt{[a+1:b]} or \texttt{[a:b-1]}).
    \item \textbf{Conditional Inversion} -- Insert or remove \texttt{not} in boolean expressions.
    \item \textbf{Range Step Modification} -- Alter the \texttt{step} parameter in \texttt{range()} calls (e.g., \texttt{range(0, n, 1)} $\leftrightarrow$ \texttt{range(0, n, 2)}).
\end{enumerate}

\subsection{Benchmark Composition Statistics}
\label{sec:AppendixBenchmarkStats}
\rev{We report detailed statistics of the 400-task evaluation set summarized in Section~\ref{sec:eval-setup}. Each task carries one or two injected corruptions (232 tasks with one, 168 with two), giving 568 corruption applications in total. We measure function lengths on AST-normalized executable function bodies after excluding comments and docstrings (Table~\ref{tab:bench_function_length}). The gold patch reverses the applied AST transformations, and we calculate token edit distance between the corrupted and canonical function bodies (Table~\ref{tab:bench_patch_size}). We observe that gold repairs are small: 50.2\% require a single token edit, 91.8\% at most two, 96.8\% at most three, and 98.0\% at most four, and every gold repair affects only one or two normalized code lines. Table~\ref{tab:bench_corruption_dist} gives the distribution over corruption families; because a family can be applied twice within a task, we report both application counts and the number of tasks containing each family. 
Table~\ref{tab:bench_bug_categories} groups the families into four descriptive semantic categories.}

\begin{table}[bt]
  \centering
  \small  \caption{\rev{Function-length distribution of the 400 tasks.}}
  \label{tab:bench_function_length}
  \setlength{\tabcolsep}{3pt}
  \begin{tabular}{@{}lrrrrrrr@{}}
    \toprule
    Measure & Mean & Med. & Q1 & Q3 & P95 & Min & Max \\
    \midrule
    Executable lines & 10.36 & 10 & 7 & 13 & 19 & 3 & 34 \\
    Python tokens & 110.18 & 102 & 77 & 134 & 199 & 32 & 364 \\
    \bottomrule
  \end{tabular}
\end{table}

\begin{table}[bt]
  \centering
  \small  \caption{\looseness=-1\rev{Minimal-patch-size distribution. The gold patch reverses the corruptions (one or two per task).}}
  \label{tab:bench_patch_size}
  \setlength{\tabcolsep}{4pt}
  \begin{tabular}{@{}lrrrr@{}}
    \toprule
    Measure & Mean & Med. & Q1 & Q3 \\
    \midrule
    Gold token edit dist. & 1.63 & 1 & 1 & 2 \\
    Normalized gold dist. & 0.017 & 0.014 & 0.010 & 0.021 \\
    Gold line edit dist. & 1.34 & 1 & 1 & 2 \\
    \bottomrule
  \end{tabular}
\end{table}

\begin{table}[bt]
  \centering
  \small  \caption{\rev{Corruption-family distribution over the 568 corruption applications. A family can be applied twice within a task, so the last column reports the number of tasks containing the family.}}
  \label{tab:bench_corruption_dist}
  \setlength{\tabcolsep}{4pt}
  \begin{tabular}{@{}lrrr@{}}
    \toprule
    Corruption family & Applic. & Share & Tasks \\
    \midrule
    Edge-case guards & 129 & 22.7\% & 77 (19.2\%) \\
    Arithmetic operators & 104 & 18.3\% & 61 (15.2\%) \\
    Comparison operators & 75 & 13.2\% & 48 (12.0\%) \\
    Numeric constants & 64 & 11.3\% & 64 (16.0\%) \\
    Range bounds & 60 & 10.6\% & 34 (8.5\%) \\
    List indexing & 43 & 7.6\% & 20 (5.0\%) \\
    Boolean constants & 40 & 7.0\% & 40 (10.0\%) \\
    Function-call substitution & 36 & 6.3\% & 36 (9.0\%) \\
    Sort order & 10 & 1.8\% & 9 (2.2\%) \\
    Slice bounds & 3 & 0.5\% & 3 (0.8\%) \\
    Accumulator initialization & 2 & 0.4\% & 2 (0.5\%) \\
    Copy removal & 2 & 0.4\% & 2 (0.5\%) \\
    \bottomrule
  \end{tabular}
\end{table}

\begin{table}[bt]
  \centering
  \small  \caption{\rev{Higher-level bug categories obtained by grouping the corruption families. Share is the fraction of the 568 corruption applications; the last column is the fraction of tasks containing the category.}}
  \label{tab:bench_bug_categories}
  \setlength{\tabcolsep}{3pt}
  \begin{tabular}{@{}>{\raggedright\arraybackslash}p{0.30\columnwidth}>{\raggedright\arraybackslash}p{0.32\columnwidth}rr@{}}
    \toprule
    Bug category & Included families & Share & Tasks \\
    \midrule
    Predicate/\newline control flow & Comparisons, edge-case guards, Boolean constants & 43.0\% & 35.5\% \\
    Computation/\newline value & Arithmetic operators, numeric constants, accumulator initialization & 29.9\% & 29.0\% \\
    Boundary/\newline iteration & Range bounds, list indexing, slice bounds & 18.7\% & 13.5\% \\
    API/data\newline semantics & Function-call substitution, sort order, copy removal & 8.5\% & 11.5\% \\
    \bottomrule
  \end{tabular}
\end{table}

\subsection{Generic and Explicit Prompts}
\label{sec:AppendixPrompt}
\looseness=-1 \rev{We give the exact prompts used in the two evaluation conditions. Both conditions share the same system prompt, which states the repair task and the signature and docstring constraints without mentioning preservation:}

{
\begin{quote}
\small\ttfamily
You are a Python Expert specializing in code analysis and debugging. When provided with a problem statement, your task is to fix the code.\\
Do not change the function signature, default arguments, or docstring. Use the docstring to understand the requirements of the function.
\end{quote}
}

\rev{We keep the signature and docstring constraint in both conditions because it represents the basic setting of code editing. The user message is shared as well, and ends with a request that carries the only text differing between the two conditions:}

{
\begin{quote}
\small\ttfamily
I am trying to implement a function with the following specifications: \{problem\_statement\}.\\[2pt]
The function I have written so far is: \{corrupted\_solution\}\\[2pt]
\textbf{\{request\}}
\end{quote}
}

\rev{followed by a fixed instruction to wrap the response in a Python code block. In the generic condition the request is:}

{
\begin{quote}
\small\ttfamily
What is wrong? Fix and complete my function.
\end{quote}
}

\rev{and in the explicit (preservation) condition it is:}

{
\begin{quote}
\small\ttfamily
What is wrong? Fix and complete my function \textbf{but keep as much of the original code as possible}.
\end{quote}
}

\looseness=-1 \rev{The intervention we measure is therefore deliberately small: a single trailing clause added to the request, shown in bold, leaving the system prompt, task description, and test suite unchanged.}

\subsection{API and Decoding Settings}
\label{sec:AppendixEvalDetails}
We ran the frontier models through the OpenAI, Anthropic, Google, and OpenRouter APIs using provider-default decoding settings (temperature \(=1\)) unless a provider required a different setting. OpenAI and Anthropic runs use batch completion APIs. Claude thinking models use a 10,000-token thinking budget and a 64,000-token maximum output length; other models use provider- and model-specific output limits through OpenRouter. For reasoning models, we use the highest available reasoning effort. In OpenRouter runs with reasoning controls, Anthropic high-reasoning models use \texttt{reasoning.max\_tokens=10000}, while other supported providers use \texttt{reasoning.effort=high}; all high-reasoning runs set \texttt{reasoning.exclude=true}. We record reasoning-token counts, available reasoning text or details, and finish reasons for audit.

\section{Validating the Edit-Fidelity Measurements}
\label{sec:AppendixValidation}
\looseness=-1 \rev{This section reports the checks supporting our use of excess Levenshtein distance and added cognitive complexity as edit-fidelity measures: a comparison against a CodeBLEU-based alternative, two human studies, an inter-model reliability check on the over-editing taxonomy, and qualitative examples.}

\subsection{Metric Sanity Check Against CodeBLEU}
\label{sec:AppendixMetricSanity}
We also \rev{compare} token-level Levenshtein distance with a CodeBLEU-based analogue of excess patch size. Because CodeBLEU is a similarity metric where higher is better, we compute the difference between the gold repair's CodeBLEU similarity to the corrupted input and the model output's similarity to that same corrupted input. Values closer to zero should therefore indicate that the model patch is about as small as the known gold repair.

The comparison gives slightly different conclusions from token edit distance. In matched-correct reasoning comparisons, CodeBLEU can favor a repair that preserves broad lexical or structural overlap while still changing unrelated code. It can also assign a model output a smaller excess score than the gold repair, implying a patch more minimal than the known reversal of the injected corruption. That behavior is difficult to interpret \rev{in} our controlled setting, where the intended local repair is known by construction.

\begin{figure*}[t]
  \centering
  \includegraphics[width=\textwidth]{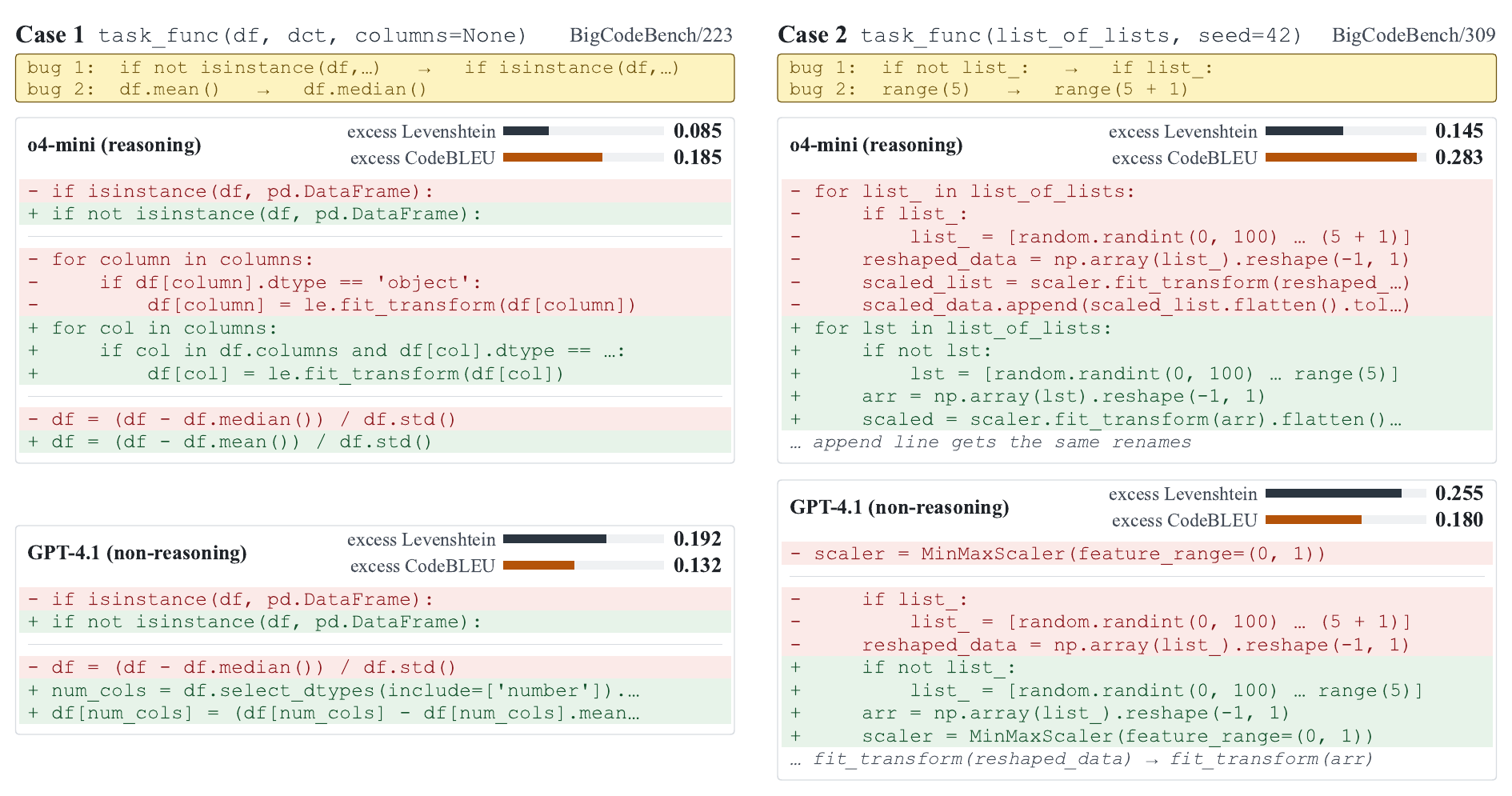}
  \caption{High-divergence examples between CodeBLEU and token-level Levenshtein distance. \rev{In two matched-correct repairs, both models fix the injected bugs (amber) and pass all tests, yet the metrics prefer opposite patches: token edit distance prefers o4-mini's smaller patch, while CodeBLEU prefers GPT-4.1's larger rewrite, because identifier renames break $n$-gram overlap more than statement rewrites that preserve surrounding text. Bars show excess over the gold repair on a shared axis (lower is better); diffs are comment-stripped function bodies. }}
  \label{fig:codebleu_levenshtein_divergence}
\end{figure*}

\rev{Figure~\ref{fig:codebleu_levenshtein_divergence} shows two matched-correct cases with high disagreement between the metrics. In both, o4-mini repairs the injected bugs with the smaller token-level patch but renames nearby variables, while GPT-4.1 rewrites entire statements; CodeBLEU rates the rewrite as the more minimal edit because long contiguous $n$-grams survive while the renames are penalized. Token edit distance better matches the intuition that a repair should disturb the implementation as little as possible.} We therefore use token-level Levenshtein distance as the main edit-size metric: it directly measures whether the model disturbed the implementation beyond reversing the injected bug, rather than rewarding broad similarity to the corrupted program.

\subsection{Human Studies}
\label{sec:AppendixHumanStudies}

\paragraph{Human validation of the edit-fidelity metrics.}
\rev{We recruited three annotators with 5--10 years of software-development experience. Each annotator evaluated the same 100 pairs of patches, where both patches in a pair repaired the same corrupted program and passed all available tests. We hid model identities and metric values, and we balanced the presentation order of Patch A and Patch B to avoid positional bias. For each pair, we asked: (1) Which patch is easier to review and understand? (2) Which patch more faithfully preserves the original implementation? \revb{The annotators are personal acquaintances of the authors from outside our research group; they} were recruited as volunteers and gave consent to share their annotations for the purpose of scientific research.}

\looseness=-1 \rev{We used the majority judgment of the three annotators as the human preference and measured its agreement with the patch preferred by each automatic metric, reporting Cohen's $\kappa$ to account for agreement expected by chance (Table~\ref{tab:human_metric_agreement}). We exclude from the denominators the cases where the human majority judgment was a tie or where no majority was reached; for added cognitive complexity, we additionally exclude pairs where both patches had the same complexity score and the metric therefore expressed no preference. We also find that the annotators agree substantially with one another on reviewability and almost perfectly on faithfulness (Table~\ref{tab:human_interannotator}; as this comparison involves three annotators, we report Fleiss' $\kappa$). Overall, we observe that excess edit distance agrees very strongly with developers' perceptions of both reviewability and faithfulness, while added cognitive complexity agrees moderately, supporting its role as a complementary measure of structural overhead.}

\begin{table}[H]
  \centering
  \small  \caption{\looseness=-1\rev{Agreement between each automatic metric and the human majority over 100 blinded patch pairs.}}
  \label{tab:human_metric_agreement}
  \setlength{\tabcolsep}{3pt}
  \begin{tabular}{@{}lcccc@{}}
    \toprule
    & \multicolumn{2}{c}{Excess edit dist.} & \multicolumn{2}{c}{Added cog.\ compl.} \\
    \cmidrule(lr){2-3} \cmidrule(lr){4-5}
    Human judgment & Agr. & $\kappa$ & Agr. & $\kappa$ \\
    \midrule
    Easier to review & 94.8\% & 0.897 & 72.7\% & 0.455 \\
    More faithful & 96.9\% & 0.939 & 69.2\% & 0.386 \\
    \bottomrule
  \end{tabular}
\end{table}

\begin{table}[H]
  \centering
  \small  \caption{\rev{Agreement among the three annotators.}}
  \label{tab:human_interannotator}
  \setlength{\tabcolsep}{3pt}
  \begin{tabular}{@{}lccc@{}}
    \toprule
    Human judgment & Pairwise agr. & Fleiss' $\kappa$ & Unanimous \\
    \midrule
    Easier to review & 84.0\% & 0.690 & 77/100 \\
    More faithful & 92.3\% & 0.850 & 89/100 \\
    \bottomrule
  \end{tabular}
\end{table}

\paragraph{Manual audit of high-excess passing repairs.}
\rev{To verify that a high excess Levenshtein distance reflects genuine over-editing rather than valid alternative repairs, we conducted a blinded manual audit of 100 sampled correct repairs with high excess distance. The annotator judged 79 repairs to contain genuinely unnecessary edits, 17 to be valid alternative local fixes, and 4 to be unclear. Among the 96 determinate cases, 82.3\% contained unnecessary edits (95\% CI 73.5\%--88.6\%), whereas 17.7\% were alternative local fixes. We conclude that high excess distance predominantly reflects over-editing.}

\subsection{Reliability of the Over-Editing Taxonomy}
\label{sec:AppendixTaxonomyValidation}
\looseness=-1 \rev{We designed and applied the taxonomy in Table~\ref{tab:overedit_issue_types} in two separate steps: we defined the categories by observable structural properties of the diffs and froze them into the annotation codebook before any labeling. The reliability question therefore concerns the labels rather than the categories.}

\rev{The shares in Table~\ref{tab:overedit_issue_types} come from an exhaustive annotation rather than a sample: GPT-5.5 (temperature $0$) labels all $530$ passing repairs with excess Levenshtein distance $\geq 0.5$ among the $10{,}000$ repairs that the five frontier models produce on the corruption-expanded set of $2{,}000$ corrupted instances. Each case presents the gold minimal patch and the model patch as unified diffs; the source model, execution status, and metric values are hidden from the annotator. Only $2.5\%$ of cases receive zero labels, and $81.5\%$ are labeled with high confidence. Table~\ref{tab:taxonomy_scopes} reports the shares under four scopes: the full population, the subset from the 400-task evaluation set, the subset excluding GPT-5.5's own repairs, and the subset excluding duplicate-fallback corruption replicates. The two dominant categories stay dominant in every scope, and no share moves by more than $12$ percentage points.}

\begin{table}[H]
  \centering
  \small  \setlength{\tabcolsep}{4pt}
  \caption{\rev{Category shares (\%) of the GPT-5.5 annotation under four scopes: all $530$ high-excess passing repairs, the 400-task evaluation subset, the subset excluding GPT-5.5's own repairs, and the subset excluding duplicate-fallback corruption replicates.}}
  \label{tab:taxonomy_scopes}
  \begin{tabular}{@{}lrrrr@{}}
    \toprule
    Category & Full & Eval set & No self & No dup. \\
    \midrule
    $n$ & 530 & 104 & 317 & 318 \\
    \midrule
    Defensive generalization & 64.2 & 65.4 & 53.0 & 59.4 \\
    Data-flow rewrite & 63.2 & 64.4 & 69.7 & 70.8 \\
    Contract drift & 34.7 & 36.5 & 29.7 & 36.5 \\
    Feature accretion & 23.6 & 23.1 & 30.0 & 25.2 \\
    Dependency fallback & 3.2 & 3.8 & 0.9 & 2.2 \\
    \bottomrule
  \end{tabular}
\end{table}

\rev{\paragraph{Labeling every passing repair.} To verify that the categories capture over-editing rather than editing per se, we extended the annotation to all $8{,}195$ passing repairs under the same frozen codebook and blinded protocol. The taxonomy is highly specific: among near-minimal repairs (excess $\leq 0$), $98.3\%$ receive zero labels and no category exceeds $1.6\%$. The share of each category rises smoothly as excess grows, rather than appearing only above the $0.5$ threshold (Table~\ref{tab:taxonomy_dose_response}). This pass also re-annotates the $530$ high-excess repairs in freshly composed batches and reproduces the labels behind Table~\ref{tab:overedit_issue_types} with per-category test--retest Cohen's $\kappa$ of $0.82$--$0.95$ (exact label-set agreement $79.1\%$, mean Jaccard $0.91$), so the reported shares are stable under re-annotation.}

\begin{table}[tb]
  \centering
  \small  \setlength{\tabcolsep}{3pt}
  \caption{\rev{Category shares (\%) among all $8{,}195$ passing repairs, over successive bins of excess Levenshtein distance $E$. Near-minimal repairs receive almost no labels, and prevalence rises smoothly with excess.}}
  \label{tab:taxonomy_dose_response}
  \begin{tabular}{@{}lrrrrr@{}}
    \toprule
    Category & $E{\leq}0$ & ${\leq}0.1$ & ${\leq}0.3$ & ${<}0.5$ & ${\geq}0.5$ \\
    \midrule
    $n$ & 2644 & 1432 & 2083 & 1506 & 530 \\
    \midrule
    Defensive generalization & 0.1 & 13.8 & 34.4 & 49.1 & 67.5 \\
    Data-flow rewrite & 0.0 & 2.4 & 11.4 & 37.1 & 66.8 \\
    Contract drift & 1.6 & 17.9 & 29.3 & 32.9 & 36.4 \\
    Feature accretion & 0.0 & 4.5 & 19.3 & 24.5 & 23.2 \\
    Dependency fallback & 0.0 & 0.2 & 0.1 & 0.4 & 3.2 \\
    \midrule
    Zero labels & 98.3 & 64.8 & 29.6 & 14.5 & 2.1 \\
    \bottomrule
  \end{tabular}
\end{table}

\rev{To assess label reliability, we re-annotated a random sample of 500 instances (100 per source model) with five independent frontier models: Claude Opus 4.6, Gemini 3.1 Pro, GPT-5.5, Grok 4.3, and DeepSeek V3.2. We gave all five models an identical blinded prompt in which we hid the source model, execution status, and metric values. Table~\ref{tab:taxonomy_agreement} shows the resulting agreement: we find that the five models show substantial to almost-perfect agreement on four of the five categories, with contract drift at moderate agreement. We take this as evidence that the categories are applied consistently across independent models rather than reflecting idiosyncrasies of GPT-5.5, the model we used for the original annotation.}

\rev{We also ran a sensitivity analysis that addresses potential self-labeling bias: when we restrict the sample to repairs \emph{not} generated by GPT-5.5, we observe no drop in agreement, so the presence of GPT-5.5 outputs among the labeled instances does not inflate agreement.}

\begin{table}[tb]
  \centering
  \small  \caption{\rev{Five-model agreement (Fleiss' $\kappa$) on the over-editing taxonomy over 500 blinded instances.}}
  \label{tab:taxonomy_agreement}
  \begin{tabular}{@{}lc@{}}
    \toprule
    Category & Fleiss' $\kappa$ \\
    \midrule
    Feature accretion & 0.90 \\
    Defensive generalization & 0.83 \\
    Dependency fallback & 0.78 \\
    Data-flow rewrite & 0.66 \\
    Contract drift & 0.40 \\
    \bottomrule
  \end{tabular}
\end{table}

\subsection{Qualitative Over-Editing Examples}
\label{sec:AppendixOvereditExamples}
Table~\ref{tab:overedit_examples} makes the categories in Table~\ref{tab:overedit_issue_types} more concrete. It shows representative high-excess repairs that still pass the tests, along with the injected corruptions, the minimal repair, and the extra behavior introduced by the model. The categories are multi-label, so a single repair can exhibit more than one kind of over-editing.

\begin{table*}[!b]
  \centering
  \scriptsize
  \setlength{\tabcolsep}{4pt}
  \renewcommand{\arraystretch}{1.08}
  \caption{Qualitative examples of passing repairs that over-edit. The table shows the local repair required for each bug and the additional behavior introduced by the model.}
  \label{tab:overedit_examples}
  \begin{tabularx}{\textwidth}{@{}>{\raggedright\arraybackslash}p{0.16\textwidth}>{\raggedright\arraybackslash}p{0.10\textwidth}>{\raggedright\arraybackslash}p{0.17\textwidth}>{\raggedright\arraybackslash}X>{\raggedright\arraybackslash}X@{}}
    \toprule
    Issue type & Model & Injected corruptions & Minimal repair & Extra model behavior \\
    \midrule
    Data-flow rewrite & DeepSeek V3.2 & Comparison operators; edge-case guards & Fix the empty-input guard and keep the simple fruit-count aggregation. & Replaces the aggregation with explicit dictionaries and loops, changing the structure of the data pipeline. \\
    Defensive generalization & GPT-5.5 & Edge-case guards; comparison operators & Change the numeric-data guard so nonempty numeric strings are accepted. & Adds JSON parsing, string/list/scalar handling, non-finite checks, exception chaining, and new plot labels. \\
    Contract drift & GPT-5.5 & Conditional inversion; arithmetic operators & Fix the interval/duration guard, loop timing, and CPU-output line index. & Buffers CPU measurements and writes one JSON array instead of preserving newline-delimited records. \\
    Feature accretion & DeepSeek V3.2 & Numeric constants; list indexing & Use 1000 samples and the correct colorbar collection index. & Builds a richer plot with a new figure, scatter points, analytic curve, mean/std lines, labels, legend, and grid. \\
    Dependency fallback & GPT-5.5 & Arithmetic operators; sort order & Increment word frequencies and remove a no-op sort argument. & Adds a hardcoded stopword fallback, exception handling, and changes the token deduplication flow. \\
    \bottomrule
  \end{tabularx}
\end{table*}

\section{Additional Evaluation Results}
\label{sec:AppendixEvalResults}
\rev{This section reports the per-model numbers behind the aggregate evaluation results in Section~\ref{sec:eval-results}, together with the open-weight prompt ablation and the statistical robustness checks for the preservation-prompt effect.}

\subsection{Per-Model Frontier Results}
\label{sec:AppendixFrontierTables}
We report the per-model frontier results behind Figure~\ref{fig:frontier_per_model_prompt_effect} in Table~\ref{tab:frontier_full}, which lists each model under both the generic and the explicit preservation prompt. The table complements the aggregate trends discussed in the main text.

\begin{table*}[tp]
  \centering
  \scriptsize
  \setlength{\tabcolsep}{4pt}
  \renewcommand{\arraystretch}{1.12}
  \caption{Frontier model performance under the generic repair prompt and the explicit preservation prompt. Higher Pass@1 and lower excess edit distance and added cognitive complexity indicate more faithful repairs; most models reduce excess edits under the explicit prompt while maintaining or improving correctness. \rev{$^\dag$The generic-prompt tests for this run were re-executed under the standard per-task limit after evaluation-harness resource starvation invalidated 112 of the 400 original test executions (Appendix~\ref{sec:AppendixRobustness}).}}
  \label{tab:frontier_full}
  \begin{tabular}{@{}l cccccc@{}}
    \toprule
    \multirow{2}{*}{Model} & \multicolumn{3}{c}{Generic prompt} & \multicolumn{3}{c}{Explicit prompt} \\
    \cmidrule(lr){2-4} \cmidrule(lr){5-7}
    & Pass@1 $\uparrow$ & Excess Lev.\ $\downarrow$ & Added CC $\downarrow$ & Pass@1 $\uparrow$ & Excess Lev.\ $\downarrow$ & Added CC $\downarrow$ \\
    \midrule
    \multicolumn{7}{l}{\textbf{Reasoning models}} \\
    Claude Opus 4.7 High \citep{anthropic2026claudeopus} & \textbf{0.912} & 0.0745 & 0.1123 & \textbf{0.930} & 0.0585 & 0.1102 \\
    Claude Sonnet 4.6 High \citep{anthropic2026_claude_sonnet46} & 0.898 & 0.0935 & 0.1309 & 0.895 & 0.0588 & 0.0726 \\
    Grok 4.3 High \citep{xai2026_grok43} & 0.863 & 0.0966 & 0.0899 & 0.878 & 0.0480 & 0.0313 \\
    GLM 5.1 High \citep{zai2026_glm51} & 0.875 & 0.1004 & 0.1314 & 0.870 & 0.0695 & 0.0920 \\
    Kimi K2.6 High \citep{moonshot2026_kimi_k26} & 0.850 & 0.1486 & 0.5353 & 0.890 & 0.0791 & 0.2949 \\
    GPT-5.5 High \citep{openai2026_gpt55} & 0.823 & 0.2986 & 1.0213 & 0.833 & 0.1589 & 0.7477 \\
    DeepSeek V3.2 High \citep{deepseekai2025deepseekv32} & 0.763 & 0.2652 & 0.9180 & 0.770 & 0.2179 & 0.8961 \\
    GPT-5.4 \citep{openai2026_gpt54_thinking_systemcard} & 0.723 & 0.3945 & 2.3125 & 0.810 & 0.2263 & 1.3500 \\
    Claude Sonnet 3.7 \citep{anthropic2025_claude37_sonnet} & 0.844 & 0.1192 & 0.4623 & 0.875 & 0.0725 & 0.3033 \\
    Claude Sonnet 4 \citep{anthropic2025_claude_sonnet4} & 0.875 & 0.1140 & 0.5013 & 0.877 & 0.0772 & 0.4261 \\
    Claude Opus 4.6 \citep{anthropic2026claudeopus46} & \textbf{0.912} & \textbf{0.0599} & 0.2000 & 0.920 & \textbf{0.0326} & 0.1125 \\
    Gemini 3.1 Pro Preview \citep{google2026_gemini31_pro} & 0.858 & 0.1448 & 0.5013 & 0.877 & 0.0765 & 0.3375 \\
    GLM 5 High \citep{zai2026_glm5} & 0.859 & 0.0991 & 0.3196 & 0.890 & 0.0523 & 0.1797 \\
    Qwen 3.6 Plus High \citep{alibaba2026_qwen36plus} & 0.858 & 0.1446 & \textbf{0.0475} & 0.873 & 0.0710 & \textbf{0.0200} \\
    Kimi K2.5 High \citep{moonshot2026_kimi_k25} & 0.835 & 0.1509 & 0.7700 & 0.882 & 0.0686 & 0.4035 \\
    DeepSeek Chat V3.1 \citep{deepseek_v31_2025} & 0.795 & 0.2320 & 0.6942 & 0.845 & 0.1633 & 0.4660 \\
    DeepSeek R1 \citep{deepseekai2025deepseekr1incentivizingreasoningcapability} & 0.820 & 0.2322 & 0.6725 & 0.782 & 0.1714 & 0.5542 \\
    Gemini 2.5 Flash High \citep{google2025_gemini25_flash} & 0.772 & 0.2578 & 1.2850 & 0.838 & 0.1517 & 0.5800 \\
    GLM 4.5 High \citep{5team2025glm45agenticreasoningcoding} & 0.810 & 0.1865 & 0.6550 & 0.835 & 0.1174 & 0.7000 \\
    GPT-5 High \citep{openai2025_gpt5_systemcard} & 0.713 & 0.4379 & 3.8321 & 0.785 & 0.2278 & 2.1754 \\
    Magistral Medium \citep{mistral2025_magistral} & 0.768 & 0.2238 & 1.2469 & 0.728 & 0.2072 & 1.2306 \\
    o4-mini High \citep{openai2025_o4_mini} & 0.770 & 0.3439 & 1.0000 & 0.838 & 0.1651 & 0.4225 \\
    Qwen3 235B A22B \citep{yang2025qwen3technicalreport} & 0.782 & 0.1823 & 0.5464 & 0.780 & 0.1574 & 0.2732 \\
    Qwen3 235B A22B Thinking 2507 \citep{qwen2025qwen3235bthinking2507} & 0.807 & 0.1882 & 0.6950 & 0.812 & 0.1331 & 0.4849 \\
    \midrule
    \multicolumn{7}{l}{\textbf{Non-reasoning models}} \\
    Claude Opus 4.7 \citep{anthropic2026claudeopus} & \textbf{0.918} & \textbf{0.0704} & \textbf{0.0845} & \textbf{0.932} & 0.0557 & 0.1635 \\
    Claude Sonnet 4.6 \citep{anthropic2026_claude_sonnet46} & 0.878 & 0.1010 & 0.2479 & 0.890 & 0.0660 & 0.1376 \\
    Grok 4.3 \citep{xai2026_grok43} & 0.878 & 0.1145 & 0.1140 & 0.870 & 0.0654 & \textbf{0.1322} \\
    GLM 5.1 \citep{zai2026_glm51} & 0.858 & 0.1087 & 0.1429 & 0.898 & 0.0655 & 0.1365 \\
    Kimi K2.6 \citep{moonshot2026_kimi_k26} & 0.840 & 0.1518 & 0.3601 & 0.885 & 0.0801 & 0.2232 \\
    GPT-5.5 \citep{openai2026_gpt55} & 0.808 & 0.2993 & 0.7895 & 0.835 & 0.1566 & 0.7635 \\
    DeepSeek V3.2 \citep{deepseekai2025deepseekv32} & 0.775 & 0.2753 & 1.2161 & 0.770 & 0.2103 & 0.8896 \\
    Qwen3-Coder 480B A35B \citep{qwen2025qwen3coder} & \rev{0.772} & 0.2196 & 0.9320 & \rev{0.830} & 0.1383 & 0.7259 \\
    Qwen3-Coder Plus$^\dag$ \citep{qwen2025qwen3coder} & 0.785 & 0.2374 & 1.0064 & 0.795 & 0.1687 & 0.9182 \\
    GPT-5.4 \citep{openai2026_gpt54_thinking_systemcard} & 0.770 & 0.3273 & 1.5625 & 0.782 & 0.2851 & 1.7200 \\
    Claude Sonnet 3.5 \citep{claude_sonnet_35_system_card} & 0.740 & 0.2778 & 1.0850 & 0.775 & 0.1982 & 0.6900 \\
    Claude Sonnet 4 & 0.830 & 0.1514 & 0.7175 & 0.848 & 0.0975 & 0.4625 \\
    Claude Opus 4.6 \citep{anthropic2026claudeopus46} & 0.900 & 0.0790 & 0.3125 & 0.912 & 0.0490 & 0.1675 \\
    Gemini 3.1 Pro Preview \citep{google2026_gemini31_pro} & 0.860 & 0.1286 & 0.3575 & 0.880 & 0.0822 & 0.3509 \\
    GLM 5 \citep{zai2026_glm5} & 0.840 & 0.0967 & 0.2350 & 0.882 & \textbf{0.0412} & 0.1425 \\
    Qwen 3.6 Plus \citep{alibaba2026_qwen36plus} & 0.870 & 0.1057 & 0.6050 & 0.880 & 0.0709 & 0.4325 \\
    Kimi K2.5 \citep{moonshot2026_kimi_k25} & 0.770 & 0.1396 & 0.6865 & 0.873 & 0.0705 & 0.2800 \\
    DeepSeek Chat V3.1 \citep{deepseek_v31_2025} & 0.802 & 0.2353 & 1.2225 & 0.840 & 0.1473 & 0.6175 \\
    DeepSeek V3 \citep{deepseekai2025deepseekv3technicalreport} & 0.800 & 0.2007 & 0.8025 & 0.785 & 0.1711 & 0.7225 \\
    Gemini 2.5 Flash \citep{google2025_gemini25_flash} & 0.759 & 0.2515 & 1.3559 & 0.800 & 0.1829 & 1.4950 \\
    GLM 4.5 \citep{5team2025glm45agenticreasoningcoding} & 0.805 & 0.1880 & 0.6800 & 0.810 & 0.1229 & 0.5125 \\
    GPT-5 Minimal \citep{openai2025_gpt5_systemcard} & 0.738 & 0.3968 & 2.8772 & 0.765 & 0.2701 & 1.9724 \\
    GPT-4.1 \citep{openai2025_gpt41} & 0.738 & 0.3104 & 1.2975 & 0.782 & 0.2136 & 0.8800 \\
    Mistral Medium \citep{mistralai_mistral_medium3} & 0.802 & 0.1319 & 0.3225 & 0.802 & 0.1194 & 0.3575 \\
    Qwen3 235B A22B 2507 \citep{yang2025qwen3technicalreport} & 0.715 & 0.3035 & 1.0225 & 0.734 & 0.2962 & 1.0902 \\
    Qwen3 235B A22B \citep{yang2025qwen3technicalreport} & 0.707 & 0.2669 & 0.7744 & 0.703 & 0.2535 & 0.7425 \\
    \bottomrule
  \end{tabular}
\end{table*}

\subsection{Open-Weight Preservation-Prompt Ablation}
\label{sec:AppendixOpenWeightPrompt}
\looseness=-1 We report the per-model open-weight prompt ablation referenced in Section~\rev{\ref{sec:eval-results}} in Table~\ref{tab:open_weight_ablation}. Prompts omit visible tests, and edit-fidelity metrics use passing repairs.

\begin{table}[H]
  \centering
  \small
  \setlength{\tabcolsep}{3.5pt}
  \caption{Open-weight model performance under generic and preservation prompts that omit visible tests. Explicit preservation improves correctness and reduces over-editing across the evaluated model families.}
  \label{tab:open_weight_ablation}
  \begin{tabular}{@{}lcccc@{}}
    \toprule
    Prompt & Pass@1 $\uparrow$ & Ex.\ Lev.\ $\downarrow$ & Add.\ CC $\downarrow$ & Over-edit $\downarrow$ \\
    \midrule
    \multicolumn{5}{@{}l}{\emph{Qwen2.5-Coder-14B-Instruct}} \\
    \quad generic & 0.803 & 0.135 & \textbf{0.146} & 0.673 \\
    \quad explicit & \textbf{0.853} & \textbf{0.099} & 0.164 & \textbf{0.540} \\
    \midrule
    \multicolumn{5}{@{}l}{\emph{Qwen2.5-Coder-32B-Instruct}} \\
    \quad generic & 0.782 & 0.155 & 0.265 & 0.732 \\
    \quad explicit & \textbf{0.820} & \textbf{0.101} & \textbf{0.207} & \textbf{0.564} \\
    \midrule
    \multicolumn{5}{@{}l}{\emph{Qwen3-Coder-30B-A3B-Instruct}} \\
    \quad generic & 0.812 & 0.159 & 0.575 & 0.674 \\
    \quad explicit & \textbf{0.848} & \textbf{0.112} & \textbf{0.510} & \textbf{0.516} \\
    \midrule
    \multicolumn{5}{@{}l}{\emph{Llama-3.1-70B-Instruct}} \\
    \quad generic & 0.767 & 0.213 & 0.336 & 0.759 \\
    \quad explicit & \textbf{0.790} & \textbf{0.153} & \textbf{0.231} & \textbf{0.665} \\
    \midrule
    \multicolumn{5}{@{}l}{\emph{Qwen3-Coder-480B-A35B-Instruct}} \\
    \quad generic & 0.772 & 0.220 & 0.932 & 0.751 \\
    \quad explicit & \textbf{0.830} & \textbf{0.138} & \textbf{0.726} & \textbf{0.584} \\
    \bottomrule
  \end{tabular}
\end{table}

\subsection{Statistical Robustness of the Preservation-Prompt Effect}
\label{sec:AppendixRobustness}

\paragraph{Paired inference for the Pass@1 gain.}
\rev{Table~\ref{tab:frontier_full} lists 50 matched generic--explicit setting pairs. One evaluation required repair: in the original generic-prompt run of Qwen3-Coder Plus, 112 of the 400 test executions (28\%) were killed by evaluation-harness resource starvation, while the same model's explicit-prompt run and every other run in the same evaluation batches were unaffected. We therefore re-executed the tests for that run's unchanged generations under the standard per-task limit; every task that originally passed still passes, and Table~\ref{tab:frontier_full} carries the corrected numbers. Across the 50 matched settings, Pass@1 increases from 81.49\% to 83.74\% ($+2.26$ percentage points), with a paired model-bootstrap 95\% confidence interval of $[+1.49, +3.05]$ points. We observe the improvement in 40 of 50 settings, and it is significant under both a two-sided Wilcoxon signed-rank test ($p = 9.88\times10^{-8}$) and a two-sided sign test ($p = 9.26\times10^{-6}$).}


\paragraph{Multiple samples per task.}
\rev{As the main evaluation uses provider-default decoding with a single sample, we verified the prompt effect under repeated sampling: for 100 randomly drawn tasks, we collected eight independent temperature-1.0 samples per model and prompt (Table~\ref{tab:repeated_sampling}). We observe that the result is consistent with the single-sample evaluation: explicit prompting reduces over-editing by 19--32\% relative and yields higher Pass@1 on all three models.}

\begin{table}[H]
  \centering
  \small  \caption{\rev{Repeated-sampling check with eight temperature-1.0 samples per task on 100 random tasks.}}
  \label{tab:repeated_sampling}
  \setlength{\tabcolsep}{4pt}
  \begin{tabular}{@{}lccc@{}}
    \toprule
    Prompt & Pass@1 $\uparrow$ & Ex.\ Lev.\ $\downarrow$ & Add.\ CC $\downarrow$ \\
    \midrule
    \multicolumn{4}{@{}l}{\emph{DeepSeek Chat V3.1}} \\
    \quad Generic & 0.766 & 0.189 & 0.878 \\
    \quad Explicit & \textbf{0.799} & \textbf{0.130} & \textbf{0.510} \\
    \midrule
    \multicolumn{4}{@{}l}{\emph{Gemini 2.5 Flash High}} \\
    \quad Generic & 0.721 & 0.240 & 1.371 \\
    \quad Explicit & \textbf{0.754} & \textbf{0.200} & \textbf{1.111} \\
    \midrule
    \multicolumn{4}{@{}l}{\emph{GPT-4.1}} \\
    \quad Generic & 0.720 & 0.308 & 1.701 \\
    \quad Explicit & \textbf{0.756} & \textbf{0.243} & \textbf{1.240} \\
    \bottomrule
  \end{tabular}
\end{table}

\paragraph{Preservation-instruction variants.}
\rev{The main evaluation tests the preservation instruction as a single fixed sentence. To check that the effect is not specific to its wording, we evaluated Qwen3-14B with three alternative preservation instructions (Table~\ref{tab:prompt_variants}): asking for the \emph{smallest-possible patch}, imposing a \emph{three-line edit budget}, and instructing the model to \emph{localize the bug, then fix it}. We find that all variants improve Pass@1 while reducing over-editing relative to the generic prompt, and that very strong constraints such as the explicit edit budget produce the smallest excess distance, at a small cost in correct repairs.}

\begin{table}[H]
  \centering
  \small  \caption{\rev{Effect of alternative preservation instructions on Qwen3-14B.}}
  \label{tab:prompt_variants}
  \setlength{\tabcolsep}{3pt}
  \begin{tabular}{@{}lccc@{}}
    \toprule
    User instruction & Pass@1 $\uparrow$ & Ex.\ Lev.\ $\downarrow$ & Add.\ CC $\downarrow$ \\
    \midrule
    Generic & 0.790 & 0.123 & 0.317 \\
    Explicit (main paper) & 0.803 & 0.074 & 0.249 \\
    Smallest-possible patch & 0.808 & 0.035 & 0.130 \\
    Three-line budget & 0.793 & \textbf{0.024} & \textbf{0.060} \\
    Localize then fix & \textbf{0.813} & 0.050 & 0.086 \\
    \bottomrule
  \end{tabular}
\end{table}


\section{Post-Training Details and Additional Analyses}
\label{sec:AppendixTraining}
\rev{This section records the training configuration for the minimal-edit experiments in Section~\ref{sec:Training}, together with the analyses that support them: auxiliary edit metrics, matched corruption counts, an RL rollout-budget ablation, and a comparison with AdaPatcher-style DPOP.}

\subsection{Minimal-Edit Training Setup}
\label{sec:AppendixTrainingDetails}
Table~\ref{tab:training} reports the in-domain and out-of-domain scores. Here, we record the run settings and held-out corruption families.

\paragraph{Optimization settings.}
We train SFT, rSFT, and DPO with LlamaFactory \citep{zheng2024llamafactory} for three epochs using a learning rate of \(10^{-5}\); for DPO, we use a preference beta of \(0.1\). We train RL with PRIME-RL \citep{primeintellect2025prime-rl} using a learning rate of \(10^{-6}\), 16 rollouts per example, and a group-mean baseline for advantage estimation. We use the execution-plus-edit-distance reward defined in Section~\ref{sec:Training}, with \(\lambda_{\mathrm{exec}}=0.1\), \(\lambda_{\mathrm{edit}}=1.0\), and \(r(M)=-0.2\) for failed or unparsable repairs. We evaluate cognitive complexity in the main-text reward ablations, while the final reward uses execution and edit distance.

\subsection{Held-Out DeepCoder Corruption Families}
\label{sec:AppendixTrainingCorruptions}
\looseness=-1 For the out-of-domain training evaluation, we apply the held-out corruption families below to DeepCoder samples. These families differ from the BigCodeBench evaluation corruptions in Appendix~\ref{sec:AppendixCorruptions}, helping us test transfer beyond a fixed corruption list.

\begin{enumerate}
    \setlength{\itemsep}{0pt}
    \setlength{\parsep}{0pt}
    \item \textbf{Min/Max Swap} -- Swap built-in functions such as \texttt{min} and \texttt{max}.
    \item \looseness=-1 \textbf{Abs Wrapping} -- Wrap an arithmetic expression in \texttt{abs()}, or unwrap an \texttt{abs()} call.
    \item \textbf{Append/Extend Toggle} -- Toggle between \texttt{list.append(x)} and \texttt{list.extend([x])}.
    \item \textbf{Enumerate Start} -- Shift the start parameter of \texttt{enumerate()}, such as implicit \texttt{0} to \texttt{1}.
    \item \textbf{Break/Continue Swap} -- Replace \texttt{break} with \texttt{continue}, or vice versa.
    \item \textbf{Dict Get Default} -- Modify or inject the default value in \texttt{dict.get()} calls.
    \item \textbf{String Case Swap} -- Toggle string case methods such as \texttt{.lower()} and \texttt{.upper()}.
    \item \textbf{Strip Variant} -- Change a strip variant, such as \texttt{.strip()} to \texttt{.rstrip()} or \texttt{.lstrip()}.
    \item \textbf{Join Separator} -- Alter the separator in \texttt{str.join()} calls.
    \item \textbf{Sorted Key Toggle} -- Add or remove a \texttt{key=len} argument in \texttt{sorted()} or \texttt{list.sort()} calls.
    \item \textbf{Set/List Cast Swap} -- Swap \texttt{set(x)} with \texttt{list(x)}, or vice versa.
    \item \textbf{Round/Int Swap} -- Replace \texttt{round(x)} with \texttt{int(x)}, or vice versa.
    \item \textbf{Comprehension Filter Removal} -- Drop one filter clause from a list, set, or dictionary comprehension.
    \item \textbf{Find/Index Swap} -- Toggle between \texttt{str.find()} and \texttt{str.index()}.
    \item \textbf{Any/All Swap} -- Replace \texttt{any(...)} with \texttt{all(...)}, or vice versa.
    \item \textbf{Zip Argument Order} -- Reverse the first two positional arguments of a \texttt{zip()} call.
    \item \textbf{Len Range Endpoint} -- Modify \texttt{range(len(x))} to \texttt{range(len(x)+1)} or \texttt{range(len(x)-1)}.
    \item \textbf{Negative Index Shift} -- Shift negative constant indices by one, such as \texttt{[-1]} to \texttt{[-2]}.
    \item \textbf{Dict Iteration Variant} -- Swap dictionary iteration methods such as \texttt{.items()}, \texttt{.keys()}, and \texttt{.values()}.
    \item \textbf{None Equality Operator} -- Toggle \texttt{None} comparisons between \texttt{==}/\texttt{!=} and \texttt{is}/\texttt{is not}.
\end{enumerate}

\subsection{Auxiliary Edit Metrics}
\label{sec:AppendixAuxMetrics}
\looseness=-1 \rev{We recomputed the out-of-domain comparison of Table~\ref{tab:training} with two auxiliary metrics, each defined analogously to excess Levenshtein distance as the model-to-buggy distance minus the reference-to-buggy distance: a raw line-level edit distance (\emph{excess line diff}) and a syntax-aware tree edit distance computed with DiffSitter\footnote{\url{https://github.com/afnanenayet/diffsitter}} (\emph{excess syntax-tree diff}), which compares parsed syntax trees and is less sensitive to surface-level differences. We observe that both auxiliary metrics produce the same model ordering as excess Levenshtein distance (Table~\ref{tab:aux_metrics}), which also correlates strongly with both at the instance level (pooled Spearman $\rho=0.909$ with line diff and $\rho=0.910$ with syntax-tree diff over the 1{,}123 correct out-of-domain repairs). We conclude that the results are not driven mainly by tokenization or formatting. We retain Levenshtein distance as the primary metric because it is more fine-grained than line diff, and added cognitive complexity because it measures added control-flow complexity; syntax-tree diff captures broader structural changes and serves as a complementary validity check.}

\begin{table}[H]
  \centering
  \small  \caption{\rev{Auxiliary excess edit metrics on correct out-of-domain repairs.}}
  \label{tab:aux_metrics}
  \begin{tabular}{@{}lcc@{}}
    \toprule
    Model & Excess line diff & Excess syntax-tree diff \\
    \midrule
    SFT & $-0.175$ & $-0.175$ \\
    rSFT & 1.721 & 1.897 \\
    DPO & 1.260 & 1.390 \\
    RL & 0.645 & 0.725 \\
    \bottomrule
  \end{tabular}
\end{table}

\subsection{Matched Corruption Counts in Training and Evaluation}
\label{sec:AppendixTrainingRobustness}
\rev{Our main setup trains with 1--10 corruptions per sample and evaluates with one or two, which could in principle confound the method comparison. We therefore repeated the full comparison with exactly one corruption per program in both the DeepCoder training set and the corrupted BigCodeBench evaluation set (Table~\ref{tab:single_corruption}). We reach the same qualitative conclusions: SFT collapses out of domain, DPO trades correctness for small edits, and RL remains the best overall combination of high Pass@1 and low excess edit distance.}

\begin{table}[H]
  \centering
  \small  \caption{\rev{Post-training comparison with exactly one corruption per program in both training and evaluation.}}
  \label{tab:single_corruption}
  \begin{tabular}{@{}lccc@{}}
    \toprule
    Model & Pass@1 $\uparrow$ & Ex.\ Lev.\ $\downarrow$ & Add.\ CC $\downarrow$ \\
    \midrule
    Untrained baseline & 0.745 & 0.089 & 0.425 \\
    SFT & 0.390 & $-0.001$ & 0.003 \\
    rSFT & \textbf{0.758} & 0.061 & 0.263 \\
    DPO & 0.640 & 0.013 & 0.005 \\
    RL & 0.753 & 0.024 & 0.113 \\
    \bottomrule
  \end{tabular}
\end{table}

\subsection{RL Rollout-Budget Ablation}
\label{sec:AppendixRolloutAblation}
\looseness=-1 \rev{rSFT and DPO use 8 generated candidates per training sample, while RL uses $K{=}16$ rollouts. To test whether the RL advantage comes from the larger sampling budget, we retrained RL with 8 and 4 rollouts (Table~\ref{tab:rollout_ablation}). We observe that decreasing the number of rollouts hurts performance only mildly, and that RL with $K{=}8$ (matching the rSFT/DPO candidate budget) still achieves a better edit-fidelity trade-off than rSFT and DPO in Table~\ref{tab:training}. We conclude that the difference in learning objective, rather than the candidate budget, is the dominant factor. Due to compute constraints, we do not perform a comprehensive hyperparameter search.}

\begin{table}[H]
  \centering
  \small  \caption{\rev{RL performance with reduced rollout budgets on out-of-domain corruptions.}}
  \label{tab:rollout_ablation}
  \begin{tabular}{@{}lccc@{}}
    \toprule
    $N_{\text{rollout}}$ & Pass@1 $\uparrow$ & Ex.\ Lev.\ $\downarrow$ & Add.\ CC $\downarrow$ \\
    \midrule
    16 & \textbf{0.782} & 0.050 & 0.185 \\
    8 & 0.763 & \textbf{0.030} & \textbf{0.115} \\
    4 & 0.755 & 0.046 & 0.166 \\
    \bottomrule
  \end{tabular}
\end{table}

\subsection{Comparison with AdaPatcher-Style DPOP}
\label{sec:AppendixAdaPatcher}
\looseness=-1 \rev{We give the full comparison against AdaPatcher-style DPO-positive (DPOP) preference learning discussed in Section~\ref{sec:Training} in Table~\ref{tab:adapatcher}; we train DPOP on our data under the DPO protocol.}

\begin{table}[H]
  \centering
  \small  \caption{\rev{Comparison with AdaPatcher-style DPOP preference learning on out-of-domain corruptions.}}
  \label{tab:adapatcher}
  \begin{tabularx}{\columnwidth}{@{}ll*{3}{>{\centering\arraybackslash}X}@{}}
    \toprule
    Method & Setup & Pass@1 $\uparrow$ & Ex.\ Lev.\ $\downarrow$ & Add.\ CC $\downarrow$ \\
    \midrule
    DPO & Full & 0.787 & 0.092 & 0.348 \\
    RL & Full & 0.782 & \textbf{0.050} & 0.185 \\
    DPOP & Full & 0.718 & 0.064 & 0.366 \\
    \midrule
    RL & LoRA $r{=}64$ & \textbf{0.797} & 0.051 & \textbf{0.160} \\
    DPOP & LoRA $r{=}64$ & 0.783 & 0.082 & 0.502 \\
    \bottomrule
  \end{tabularx}
\end{table}

\section{Responsible Research}
\label{sec:AppendixResponsibleResearch}

\subsection{Potential Risks}
\looseness=-1 We study code-repair systems, so this work may indirectly improve automated software modification. The main risk is over-trust in model-generated edits: a repair can pass tests while changing unrelated behavior, increasing review burden or hiding regressions outside the test suite. We focus our experiments on public benchmark programs and synthetic, reversible corruptions, and we do not study vulnerability discovery, exploit generation, or deployment against production repositories. Throughout the paper, we advocate evaluating edit fidelity alongside functional correctness and keeping model-produced repairs subject to human review.

\subsection{Licenses and Intended Use of Scientific Artifacts}
\label{sec:AppendixScientificArtifacts}

We use third-party artifacts for research evaluation or training in the manner described by their documentation. BigCodeBench \citep{zhuo2025bigcodebenchbenchmarkingcodegeneration}, released under Apache 2.0, provides the 400 Python repair tasks, reference solutions, and tests used in the main evaluation; we inject local AST-level corruptions into the reference solutions and evaluate repairs with the original executable tests. The public DeepCoder preview resources \citep{luo2025deepcoder}, released under MIT terms, provide coding samples for the minimal-edit training experiments, where we create corrupted training and evaluation instances and keep only corrupted programs that fail tests. LiveCodeBench \citep{jain2024livecodebenchholisticcontaminationfree}, also released under MIT terms, serves as a held-out broader coding benchmark; we report LiveCodeBench v6 deltas to check whether minimal-edit training harms general coding ability.

\looseness=-1 For open-weight inference and fine-tuning, we use Qwen coding and Qwen3 models \citep{hui2024qwen25coder,yang2025qwen3technicalreport}. The Qwen releases used here are distributed under Apache 2.0 except where an individual model card states otherwise; we evaluate Qwen2.5-Coder and Qwen3-Coder variants and fine-tune Qwen3 4B/14B models for minimal editing. We also use Llama-3.1-70B-Instruct \citep{meta2024llama31}, governed by the Llama 3.1 Community License, as an open-weight prompt-ablation baseline through hosted inference. For training infrastructure, we use LlamaFactory \citep{zheng2024llamafactory} for SFT, rSFT, and DPO, and PRIME-RL \citep{primeintellect2025prime-rl} for GRPO-style reinforcement learning; both frameworks are released under Apache 2.0. For closed API models, we use provider-hosted inference under the corresponding API terms and report aggregate measurements and qualitative examples.

\subsection{Data Composition and Privacy}
\looseness=-1 We build the primary evaluation set from 400 BigCodeBench Python tasks, each with \rev{one or two} injected corruptions, and use a \(400\times5\) variant-expanded set aggregated over \(10{,}000\) model-case evaluations for the corruption analysis. We work with public code benchmarks and synthetic program transformations; we do not use private repositories, personally identifying information, or intentionally offensive content. \rev{The only human data we collect are the expert judgments in the annotation studies of Appendix~\ref{sec:AppendixHumanStudies}, where annotators compared pairs of model patches.}

\subsection{Compute}
We conduct all experiments using NVIDIA A100 GPUs. Each run of 4B RL \rev{for} 100 steps takes around 3 hours on 8 GPUs. 

\subsection{AI Assistant Use}
\looseness=-1 We used LLMs for limited research support, including organizing experiment code and polishing writing, plots, and tables. We did not use them for study design, artifact selection, experiment execution, result verification, \rev{or} final claims.


\end{document}